\documentclass[10pt]{article}
\usepackage{graphicx}
\usepackage{epsfig}
\usepackage{rotating}
\usepackage{amsmath}
\usepackage{amsfonts}
\usepackage{amssymb}
\usepackage{url}
\usepackage{hyperref}
\usepackage[footnotesize]{caption2}
\newcommand{\ba}{\begin{array}{c}}
\newcommand{\ea}{\end{array}}
\newcommand{\baz}{\begin{array}{cc}}
\def\ba{\begin{eqnarray}}
\def\ea{\end{eqnarray}}
\def\br{\begin{array}}
\def\er{\end{array}}
\def\be{\begin{equation}}
\def\ee{\end{equation}}

\def\da1{d\alpha_1\over dt}
\def\da2{d\alpha_2\over dt}

\def\cj{{\cal J}}

\newcommand{\bea}{\begin{eqnarray}}
\newcommand{\eea}{\end{eqnarray}}
\newcommand{\bmt}{\begin{pmatrix}}
\newcommand{\emt}{\end{pmatrix}}

\begin{document}
\noindent{\large\bf{New mechanism for Type-II seesaw dominance in SO(10) with low-mass $Z^{\prime}$, RH neutrinos, 
and verifiable LFV, LNV and proton decay}}
\author{Bidyut Prava Nayak\inst{1} \and Mina Ketan Parida\inst{2}}
%
%
\begin{center}
{\bf Bidyut Prava Nayak ${}^{\dag}$ and M. K. Parida ${}^{*}$}\\

{{{\em Centre of Excellence in Theoretical and Mathematical
~Sciences\\
{\rm SOA} University,
 Khandagiri Square, Bhubaneswar  751030, India}}}\\ 
{\sl${}^{*}$email:parida.minaketan@gmail.com}\\
{\sl${}^{\dag}$email:bidyutprava25@gmail.com}\\
\end{center}

\date{Received: date / Revised version: date}
%
\abstract{The dominance of type-II seesaw mechanism for neutrino masses has
attracted considerable attention because of a number of advantages.
We show a novel approach to achieve Type-II seesaw
dominance in non-supersymmetric $SO(10)$ grand
unification where a low mass $Z^{\prime}$ boson and specific patterns
of right-handed neutrino masses are predicted within
the accessible energy range of the Large Hadron Collider.   
 In spite of the high value
of the seesaw scale, $M_{\Delta_L} \simeq 10^8-10^9$ GeV, the model predicts new
dominant contributions to neutrino-less double beta decay 
 in the $W_L-W_L$
channel close to the current experimental limits via exchanges of heavier singlet fermions used as essential
ingredients of this model even when the light active neutrino masses
are normally hierarchical or invertedly hierarchical. We obtain upper bounds on the lightest sterile neutrino mass $m_s\lesssim 3.0$ GeV, $2.0$ GeV, and
 $0.7$ GeV for normally hierarchical, invertedly hierarchical and quasi-degenerate patterns of light neutrino masses, respectively. 
 The underlying non-unitarity effects lead to lepton flavor violating decay branching
ratios  within the reach of ongoing or planned 
 experiments and the leptonic CP-violation
parameter nearly two order larger than the quark sector. Some of the predicted
values on proton
lifetime  for $p\to e^+\pi^0$ are found to be within the currently
accessible search limits. Other aspects of model applications
including leptogenesis  etc. are
briefly indicated.}

%

%
\section{\small\bf INTRODUCTION}
Experimental evidences on tiny neutrino masses and their large mixings have attracted
considerable attention as physics beyond the standard model (SM)
leading to different  mechanisms for neutrino mass generation. Most of
these models are based upon the underlying assumption that neutrinos
are Majorana fermions that may manifest in the detection of events in
neutrino-less double beta ($0\nu\beta\beta$)
decay experiments on which a number of investigations are in progress
\cite{bbexpt1,bbexpt2,bbexpt3}. 
Theories of neutrino masses and mixings are placed on a much stronger
footing if they originate from left-right symmetric (LRS) \cite{ps,rnmpati}    grand unified
theories such as SO(10) where, besides grand unification of three
forces of nature, P ($=$Parity) and CP-violations have spontaneous-breaking
origins, the fermion masses of all the three generations are adequately
fitted \cite{baburnm93}, all the $15$ fermions plus the right-handed neutrino ($N$) are
unified into a  single spinorial representation ${\bf 16}$ and
the canonical ($\equiv$ type-I )  seesaw formula for neutrino masses
is predicted by the theory. More recently non-SUSY $SO(10)$ origin of
cold dark matter has been also suggested \cite{kadastik:2009}. 
Although type-I seesaw formula was also proposed by using extensions of
the SM  \cite{type-I,valle:T1}, it is well known that this  was advanced  even much before the atmospheric neutrino
oscillation data \cite{kajita98} and it is interesting to note that Gell-Mann, Ramond
and Slansky  had used the left-right symmetric  SO(10) theory and its Higgs representations ${10}_H ,{126}_H$ to
derive it.
 A special feature of left-right (LR) gauge theories and SO(10) grand unification is that the canonical seesaw formula for neutrino masses 
is always accompanied by type-II seesaw  formula \cite{type-II} for Majorana neutrino mass matrix     
\begin{equation}                     
         \mathcal{ M_\nu}=m^{II}_\nu+m^I_\nu ,\\ 
         \end{equation}
         \bea
          m^I_\nu = -M_D\frac{1}{M_N}M_D^T ,
          \eea
          \bea
          m^{II}_\nu = fv_L           
          \eea
         \label{one-twoseesaw}
where $M_D(M_N)$ is Dirac (RH-Majorana) neutrino mass, $v_L$ is the
induced vacuum expectation value (VEV) of the left-handed (LH) triplet
$\Delta_L$, and $f$ is the Yukawa coupling of the triplet. Normally,
because of the underlying quark-lepton symmetry in SO(10), $M_D$ is of
the same order as $M_u$, the up-quark mass matrix. Then the neutrino
oscillation data forces the canonical seesaw scale to be large, $M_N\geq 10^{11}$ GeV. Similarly the type-II seesaw scale is also large.
With such high seesaw scales, these two mechanisms in SO(10) can not be directly
verified at low energies or by the Large Hadron Collider (LHC) except for
the indirect signature through
the light active neutrino mediated  $0\nu \beta\beta$ decay, and
possibly leptogenesis.\\ 
It is
well known that the theoretical predictions of branching ratios for LFV decays
such as $\mu\to e \gamma$, $\tau \to \mu \gamma$, $\tau \to e
\gamma$, and $\mu \to e{\bar e}e$ closer to their experimental limits
are generic features of SUSY GUTs even with high seesaw
scales but, in non-SUSY models with such seesaw scales, they are far
below the experimental limits.  
Recently they have been also predicted to be experimentally accessible 
along with low-mass $W_R,Z_R$ bosons
through TeV scale gauged inverse seesaw mechanism \cite{psbrnm} in
SUSY SO(10). In
the absence of any evidence of supersymmetry so far, alternative  non-SUSY SO(10) models have been
found with predictions of substantial LFV decays and TeV scale
$Z^{\prime}$ 
bosons. 
Although two-step
 breakings of LR gauge theory was embedded earlier in non-SUSY GUTs with
 low-mass $Z'$ \cite{mkpzp}, its successful compliance with neutrino
 oscillation data has been possible in the context of inverse seesaw
 mechanism and predictions of LFV decays \cite{ramlalmkp}, or with the predictions of low-mass
$W_R,Z_R$ bosons, LFV decays, observable neutron oscillations, and
dominant LNV decay via extended seesaw
mechanism \cite{app}.
Possibility of LHC accessible
low-mass $Z^{\prime}$ has been also investigated recently in the context of
  heterotic string models \cite{faraggi}. 
 Another attractive aspect of non-SUSY SO(10) is rare kaon
decay and neutron-antineutron oscillation which has been discussed  in a
recent work with
inverse seesaw mechanism for light neutrino masses and TeV scale
$Z^{\prime}$ bosons but having much larger $W_R$ mass not accessible to LHC \cite{pas}. The
viability of 
the model of ref.\cite{psbrnm} depends on the discovery of TeV scale SUSY
 , TeV scale $W_R,Z_R$ bosons, and TeV scale pseudo-Dirac neutrinos. The viability of the non-SUSY model
 of ref.\cite{ramlalmkp} depends on
the discovery of  TeV scale low-mass $Z_R$ boson and heavy
 pseudo Dirac neutrinos in the range $100-1200$  GeV; both types of
 models predict proton lifetime within the Super-K search limit.
The falsifiability of the non-SUSY model of ref.\cite{pas} depends
upon any one of the following  predicted observables: TeV scale $Z_R$ boson,
dominant neutrino-less double beta decay, heavy Majorana type sterile
 and right-handed neutrinos, neutron oscillation, and rare kaon decays.
Whereas the neutrino mass generation mechanism in all these models is
through gauged inverse seesaw mechanism, our main thrust in the
present work is type-II seesaw.                      
 A key ansatz to resolve the issue of large mixing in the neutrino sector
 and small mixing in the quark sector has been suggested to be through type-II
seesaw  dominance \cite{mpr} via renormalisation group evolution
of quasi-degenerate neutrino masses that holds in supersymmetric
quark-lepton unified theories \cite{ps} or SO(10) and for large values of $\tan\beta$ which
represents the ratio of vacuum expectation values (VEVs) of up-type
and down type Higgs doublets. In an interesting approach to understand
neutrino mixing in SUSY theories, it has been shown \cite{goran1} that the maximality
of atmospheric neutrino mixing is an automatic cosnsequence of type-II
seesaw dominance and $b-\tau$ unification that does not require
quasi-degeneracy of the associated neutrino masses. A number of
consequences of this approach have been explored to
explain all the fermion masses and mixings by utilising type-II seesaw,
or a combination of both type-I and type-II seesaw \cite{rnm2,misc} through SUSY
SO(10). 
As a further interesting property of type-II seesaw dominance, it has been 
recently shown \cite{alta} without using any flavor
symmetry  that the well
known tri-bimaximal mixing pattern for neutrino mixings is simply a
consequence of rotation in the flavor space. Although several models of Type-II
seesaw dominance in SUSY SO(10) have been investigated, precision
gauge coupling unification is  distorted in most cases\footnote{ A brief review of different SUSY
  $SO(10)$ models requiring type-II seesaw, or an admixture of type-I
  and type-II for fitting fermion masses is given in
 ref. \cite{alta}.
and a brief review of distortion occuring to
  precision gauge coupling unification is given in ref. \cite{rnmmkp11}.}.
 All the charged fermion mass fittings in the conventional one-step breaking of SUSY
GUTs including fits to the neutrino oscillation data 
require the left-handed triplet to be lighter than the type-I seesaw
scale. The gauge coupling evolutions being sensitive to the quantum
numbers of the LH triplet $\Delta_L(3, -2, 1)$ under SM gauge
group, tend to misalign the precision unification in the minimal
scenario achieved without the lighter triplet.\\
 Two kinds of SO(10) models have been suggested for
ensuring precision gauge
coupling unification in the presence of type-II seesaw dominance.
  In the first type of SUSY model \cite{rnm3}, SO(10)
breaks at a very high scale $M_{U}\ge 10^{17}$ GeV to SUSY SU(5)
which further breaks to the minimal supersymmeric standard model (MSSM) at the usual SUSY GUT scale $M_{U}\sim
2\times 10^{16}$ GeV. Type-II seesaw dominance is achieved by fine
tuning the mass of the 
full $SU(5)$ multiplet ${15}_H$ containing the $\Delta_L (3,-2,1)$ to remain at
the desired type-II scale $M_{\Delta_L}=10^{11}-10^{13}$ GeV. Since
the full multiplet ${15}_H$ is at
the intermediate scale, although the evolutions of the three gauge
couplings of the MSSM gauge group deflect from their original paths 
 for $\mu > M_{\Delta_L}$,  
they converge exactly at the same scale $M_U$ as the MSSM unification
scale but with
a slightly larger value of the GUT coupling leading to a marginal
reduction of proton-lifetime prediction compared to SUSY SU(5).
In the second class of models applicable to a non-SUSY  or 
split-SUSY case \cite{rnmmkp11}, the grand unification group SO(10)  breaks directly to the SM gauge symmetry at the
GUT-scale $M_{U}\sim
2\times 10^{16}$ GeV and by tuning the full SU(5) scalar multiplet
${15}_H$ to have degenerate masses at  $ M_{\Delta_L}=
10^{11}-10^{13}$ GeV, the type-II seesaw dominance is achieved. The question of precision unification 
is answered in this model  by pulling out all the super-partner 
scalar components 
of the MSSM  but by keeping all the fermionic superpartners and the
two Higgs doublets near the
TeV scale. In the non-SUSY case the TeV scale fermions can be also
equivalently replaced by complex scalars carrying the same quantum
numbers. 
The proton lifetime prediction is $\tau_P(p\to e^+\pi^0) \simeq
10^{35}$ Yrs. in this model.\\    
In the context of LR gauge theory, type-II seesaw mechanism was
originally proposed with manifest left right symmetric gauge group $SU(2)_L\times SU(2)_R\times
U(1)_{B-L}\times SU(3)_C\times D$ ($g_{2L}=g_{2R}$) ($\equiv
G_{2213D}$) where both the left- and the right-handed triplets are
allowed to have the same mass scale as the LR symmetry breaking (or
the Parity breaking ) scale \cite{ms:1981}. With the emergence of D-Parity and its
breaking leading to decoupling of Parity and $SU(2)_R$ breakings \cite{cmp}, a new
class of asymmetric LR gauge group also emerged:
$SU(2)_L\times SU(2)_R\times
U(1)_{B-L}\times SU(3)_C$ ($g_{2L} \neq g_{2R}$) ($\equiv
G_{2213}$) where the left-handed triplet acquired larger mass than
the RH triplet leading to the type-I seesaw dominance and suppression
of type-II seesaw in $SO(10)$ \cite{cm}. It is possible to accommodate
both types of intermediate symmetries in non-SUSY SO(10) but these
models make negligible predictions for branching ratios of charged
LFV processes and they leave no other experimental signatures to be
verifiable at low or LHC energies except $0\nu\beta\beta$ decay.\\   
The purpose of this work is to show that in a class of 
 models descending from non-SUSY SO(10) or from Pati-Salam gauge symmetrty,
 type-II seesaw dominance at intermediate scales ($M_{\Delta}\simeq
10^8-10^9$ GeV) but with $M_N\sim O(1)- O(10)$ TeV can be realised
 by cancellation of the type-I seesaw contribution along with the
 prediction of a
 $Z^{\prime}$ boson at $\sim O(1)-O(10)$ TeV scale accessible to the large Hadron Collider
 (LHC) where $U(1)_R\times U(1)_{B-L}$ breaks spontaneously to
 $U(1)_Y$ through the VEV of the  RH
 triplet component of Higgs scalar
 contained in ${126}_H$ that carries $B-L= -2$. 

Although two-step
 breakings of LR gauge theory was embedded earlier in non-SUSY GUTs with
 low-mass $Z'$ \cite{mkpzp}, its successful compliance with neutrino
 oscillation data has been possible in the context of inverse seesaw
 mechanism \cite{ramlalmkp}.

We also discuss how the type-II seesaw contribution dominates over the
 linear seesaw formula. Whereas in all
 previous Type-II seesaw dominance models in SO(10), the RH Majorana neutrino
 masses have been very large and inaccessible for accelerator
 energies, the present model predicts these masses in the LHC
 accessible range. In
 spite of large values of the $W_R$ boson and the
 doubly charged Higgs boson $\Delta_L^{++}, \Delta_R^{++}$ 
 masses, it is quite interesting to note that the model predicts a
 new observable contribution to $0\nu\beta\beta$ decay in the
 $W_L-W_L$ channel. The key ingredients to achieve type-II seesaw
 dominance by complete suppression of type-I seesaw contribution are
 addition of one SO(10) singlet fermion  per generation ($S_i,i=1,2,
 3$) and
 utilization  of the additional Higgs representation ${16}_H$ to
 generate the $N-S$ mixing term in the Lagrangian through Higgs-Yukawa
 interaction.    
The underlying leptonic non-unitarity effects lead to substantial LFV decay
 branching ratios and leptonic CP-violation accessible to ongoing
 search experiments. We derive a new
 formula for the half-life of $0\nu\beta\beta$ decay as a function of the fermion singlet masses and extract 
 lower bound on the lightest sterile neutrino mass from the existing
 experimental lower bounds on
 the half-life of
 different experimental groups. For certain regions of parameter space
 of the model, we also find the proton lifetime for
 $p\to e^+\pi^0$ to be accessible to ongoing or planned experiments. \\ 
Compared to earlier existing SO(10) based type-II seesaw dominant models whose
RH neutrino masses are in the inaccessible range and new gauge bosons are in
the mass range $10^{15}-10^{17}$ GeV, the present model predictions on LHC
scale $Z^{\prime}$, light and heavy Majorana type sterile neutrinos, RH Majorana
neutrino masses in the range $\simeq {100-10000}$ GeV accessible to LHC in
the $W_L-W_L$ channel through dilepton production, the LFV branching ratios
closer to experimental limits, and dominant $0\nu\beta\beta$ decay amplitudes
caused by sterile neutrino exchanges provide a rich testing ground for new physics
signatures.\\
This paper is organized as follows. In Sec.2. we give an outline
of the model and discuss  gauge coupling unification along with
 proton lifetime predictions. In Sec.3  we derive type-II
 seesaw dominance formula and show how the model predicts RH neutrino
 masses from fits to the neutrino oscillation data.  
In Sec.4 we discuss the derivation of  Dirac neutrino mass matrix from
the GUT scale fit to fermion masses. 
 In Sec.5 we discuss predictions on lepton flavor violation and
 leptonic CP violation due to the underlying non-unitarity effects.
In Sec.6 we discuss briefly analytic derivation of amplitudes on lepton number 
violation. In Sec.7 we discuss predictions on effective mass
parameters and half life for $0\nu\beta\beta$  
where we also obtain the singlet fermion mass bounds. We also indicate
very briefly some plausible model applications including effects on electroweak precision
observables, $Z-Z'$ mixings, dilepton production, and leptogenesis in Sec.8. We summarize and conclude our
results in Sec.9.
\section{\small\bf UNIFICATION WITH TeV SCALE
  $Z^{\prime}$}\label{Sec.2}
In this section we devise two symmetry breaking chains of non-SUSY $SO(10)$
theory, one with LR symmetric gauge theory with unbroken D-Parity and
another without D-Parity at the intermediate scale. In the subsequent
sections we will compare the ability of the two models to accommodate
type-II seesaw dominance to distinguish one model from the other. As
necessary requirements, we introduce one $SO(10)$-singlet per
generation ($S_i,i=1,2,3$) and Higgs representations ${126}_H$ and
${16}_H$ in both the models .       
\subsection{Models from $SO(10)$ symmetry breaking}
Different steps of symmetry breaking is given below for the following
two models:
\par\noindent{\bf {\underline {Model-I}}}
\ba
SO(10) &
\stackrel {(M_U=M_P)}
{\longrightarrow}      
& SU(2)_L \times
SU(2)_R \times U(1)_{B-L}\times  SU(3)_C~~[G_{2213}] \nonumber \\ 
&
\stackrel {(M_R^+)}
{\longrightarrow}& SU(2)_L \times
U(1)_R \times U(1)_{B-L}\times  SU(3)_C~~[G_{2113}] \nonumber \\
&
\stackrel {(M_R^0)}
{\longrightarrow}& SU(2)_L \times
U(1)_Y\times  SU(3)_C~~[\rm SM] \nonumber\\
&
\stackrel {(M_Z)}{\longrightarrow}&SU(3)_C \times U(1)_Q, \nonumber
\ea
\par\noindent{\bf {\underline {Model-II}}}
\ba
SO(10) &
\stackrel {(M_U)}
{\longrightarrow}      
& SU(2)_L \times
SU(2)_R \times U(1)_{B-L}\times SU(3)_C[G_{2213D}]\nonumber \\ 
&
\stackrel {(M_R^+=M_P)}
{\longrightarrow}& SU(2)_L \times
U(1)_R \times U(1)_{B-L}\times  SU(3)_C~~[G_{2113}] \nonumber \\
&
\stackrel {(M_R^0)}
{\longrightarrow}& SU(2)_L \times
U(1)_Y\times  SU(3)_C~~[\rm SM] \nonumber\\
&
\stackrel {(M_Z)}{\longrightarrow}&SU(3)_C \times U(1)_Q. \nonumber
\ea
In Model-II, $SU(2)_L \times
SU(2)_R \times U(1)_{B-L}\times SU(3)_C\times D~~[\equiv G_{2213D}](g_{2L} =
g_{2R})$ is obtained by breaking the GUT-symmetry and by giving vacuum
expectation value (VEV) to the D-Parity even singlet $(1,1,0,1)\subset
(1,1,15)\subset
{210}_H$ \cite{cmp,cmgmp} where the first, second, and the third set of
quantum numbers of the scalar components are under $G_{2213P}$, the
Pati-Salam symmetry $G_{224}$, and $SO(10)$, respectively. As a
result, the Higgs sector is symmetric below $\mu=M_U$ leading to equality between 
the gauge couplings $g_{2L}(M_R^+)$ and $g_{2R}(M_R^+)$ . In this case
the LR discrete symmetry ($\equiv$ Parity) survives down to the intermediate scale ,$M_{R^+}=M_P$. The second step of symmetry breaking 
is implemented by assigning VEV to the neutral component of the right-handed (RH) Higgs triplet $\sigma_R(1,3,0,1)\subset {45}_H$ that carries $B-L=0$.       
The third step of breaking to SM is carried out by assigning VEV of  
 ${\mathcal O}(5-10)$ TeV to the $G_{2113}$ component   $\Delta^0_R(1,1,-2,1)$ contained in the RH triplet $\Delta_R(1,3,-2,1)\subset {126}_H$ carrying $B-L=-2$.
  This is responsible for RH Majorana neutrino mass generation $M_N=fV_R$ where
$V_R=\langle\Delta_R^0\rangle$ and $f$ is the Yukawa coupling of ${126}^{\dagger}$ to $SO(10)$ spinorial fermionic representation :$f{\bf 16.16. 126^{\dagger}_H}$.
We introduce $SO(10)$ invariant $N-S$  mixing mass via the Yukawa interaction $y_{\chi}{\bf
  16.1.16_{H}^{\dagger}}$ and obtain the mixing mass $M=y_{\chi}V_{\chi}$ where $V_{\chi}=\langle\chi_R^0\rangle$ by noting that  under $G_{2113}$ the submultiplet $\chi_R^0(1,1/2,-1,1)$ is contained in the $G_{2213}$ doublet  $\chi_R(1,2,-1,1)\subset 16_H$. The symmetry breaking in the last step is implemented through the SM Higgs doublet contained in 
the bidoublet $\phi(2,2,0,1) \subset {10}_H$ of SO(10). This is the
minimal Higgs structure of the model, although we will utilise two different Higgs doublets $\phi_u \subset 
10_{H_1}$ and $\phi_d \subset 10_{H_2}$ for fermion mass fits.
In Model-I, the GUT symmetry breaks to LR gauge symmetry 
$G_{2213}(g_{2L}\neq g_{2R})$ in such a way that the D-parity breaks
at the GUT scale and is decoupled from $SU(2)_R$ breaking that occurs
at the intermediate scale. This is achieved by giving GUT scale VEV to
the D-parity odd singlet-scalar component in $(1,1,0,1)_H\subset
(1,1,15)_H \subset {45}_H$ where the first, second , and third
submultiplets are under $G_{2213}$, the Pati-Salam symmetry $G_{224}$,
and $SO(10)$, respectively. In this case by adopting the D-Parity
breaking mechanism \cite{cmp} in $SO(10)$,  normally the LH triplet
component $\Delta_L(3,1,-2,1) \subset {126}_H$ and the LH doublet
component $\chi_L(2,1,-1,1)\subset {16}_H$ acquire  masses at the GUT
scale while the RH triplet and RH doublet components, $\Delta_R(1, 3, -2,1) \subset {126}_H$ 
$\chi_R(1, 2 ,-1,1)\subset {16}_H$, can be made much lighter. 
We have noted that in the presence of color octet at lower scales,
found to be necessary in this Model-I as well as in Model-II, precision gauge coupling is
achieved  even if the the parameters of the Higgs
potential are tuned so as to have the LH triplet mass at intermediate
scale, $M_{\Delta_L} \simeq 10^8-10^9$ ~GeV. The presence of
$\Delta_L(3,1,-2,1)$ at the intermediate scale plays a crucial role in
achieving Type-II seesaw dominance as would be explained in the
following section. The necessary presence of lighter LH triplets in
GUTs  with or without
vanishing $B-L$ value for physically appealing predictions was pointed
out earlier in achieving observable matter anti-matter oscillations
\cite{mkp}, in the context of low-scale leptogenesis 
\cite{mkparc10}, and type-II seesaw dominance in SUSY, non-SUSY and
split-SUSY models \cite{rnm3,rnmmkp11}, and  also for TeV scale LR
gauge theory originating from SUSY $SO(10)$ grand
unification\cite{psbrnm}.
\subsection{Renormalization group solutions to mass scales}
In this section while safeguarding precise unification of gauge
couplings at the GUT scale, we discuss allowed solutions of
renormalization group equations (RGEs) for the mass scales  $M_U, M_{R^+}$,
and $M_{R^0}$ as a function of the mass $M_C$ of the lighter color
octet $C_8(1,1,0,8)\subset {45}_H$.
The Higgs scalars contributing to RG evolutions are presented in Table 
\ref{tab:particle} for Model I. In Model II, in addition to the Higgs
scalars shown in  Table \ref{tab:particle}, 
the masses of the left handed scalars $\chi_L(2,1,-1,1)$ and
 $\sigma_L(3,1,0,1)$ are naturally constrained to be at $\mu=M_R^+=M_P=$ the parity violation scale.

Higgs scalars 
\begin{table}[htbp]
\centering
\begin{tabular}{|l|l|c|}
\hline
Mass scale $(\mu)$&Symmetry& Higgs scalars (Model-I)\\
\hline
 $M_Z-M_R^{0}$&$G_{213}$&$\Phi(2,1,1)$\\
\hline
 $M_R^0-M_R^{+}$&$G_{2113}$&${\small \begin{array}{l}
\Phi_1(2,1,0,1), \Phi_2(2,1,0,1), \\[2mm] \chi_R(1,1/2,-1,1),
\Delta_R(1,1,-2,1), \\[2mm] C_8(1,0,0,8) \end{array}}$   \\
\hline
 $M_R^+-M_U$&$G_{2213}$&${\small \begin{array}{l}\Phi_1(2,2,0,1), \Phi_2(2,2,0,1)),\\[2mm] \chi_R(1,2,-1,1),
 \Delta_R(1,3,-2,1), \\[2mm] \Delta_L(3,1,-2,1), C_8(1,0,0,8), \\[2mm]\sigma_R(1,3,0,1)\end{array}}$\\
\hline
\end{tabular}
\caption{Higgs scalars and their quantum numbers used in Model-I in the respective ranges of mass scales. The mass of colour octet $C_8(1,0,0,8)$ has
been used in the range $10^4-10^{11}$ GeV contributing to the variation of
predicted proton lifetime discussed below in Sec.2.3. In Model-II, in addition to these scalars,
 the masses of left handed scalars $\chi_L(2,1,-1,1)$ and
 $\sigma_L(3,1,0,1)$ are constrained to be at $\mu=M_R^+=M_P=$ the parity violation scale.}
\label{tab:particle}
\end{table}

The renormalisation group (RG) coefficients for
the minimal cases have been given in Appendix A to which those due to
the color octet scalar in both models and the LH triplet $\Delta_L$
in Model-I in their suitable ranges of the running scale have been added. \\   
\par\noindent{\bf Model-I:}\\
 As shown in Table \ref{tab:gauge-gut1}{ for Model-I, with $M_{\Delta_L}= 10^8$ GeV  the
 $G_{2213}$ symmetry is found to survive down to $M_{R^+}=
 (10^{8}-10^{10})$ GeV  with larger or smaller unification scale
 depending upon the color octet mass. In particular we note one set of solutions,
\ba
M_{R^0}&=&10~{\rm TeV}, ~M_{R^+}=10^{9.7} {\rm GeV},~M_{U}=10^{15.62}
{\rm GeV}, \nonumber\\
M_{\Delta_L}&=&10^8~{\rm GeV}, ~M_C=10^{10.9} {\rm GeV}.
\label{mod1sol}
\ea
As explained in the following sections, this set of solutions are found to be attractive both from the prospects of 
achieving type-II seesaw dominance and detecting proton decay at Hyper-Kamiokande.   
 with $M_U= 6.5\times 10^{15}$ GeV when the color octet mass is at
 $M_C\sim 10^{11}$ GeV. As discussed below the proton lifetime in this
 case is closer to the current experimental limit. With allowed values
 of $M_{R^0}= (5-10)$ TeV, this model also predicts $M_{Z^{\prime}}
 \simeq (1.2-3.5)$ TeV in the accessible range of the Large Hadron
 Collider. As discussed in the following Sec.\ref{Sec.3}, because of the
 low mass of the $Z^{\prime}$ boson 
 associated with TeV scale VEV of $V_R$, the type-II seesaw mechanism
 predicts RH neutrino masses which can be testified at the LHC or future high
 energy accelerators.  
 \begin{table}[htbp]
\centering
\begin{tabular}{|l|l|c|c|c|c|}
\hline
$M_R^0$&$M_C$&$M^+_R$&$M_G$&$\alpha_G^{-1}$ & $\tau_p$\\
(TeV)&(GeV)&(GeV)&(GeV)&&(Yrs.)\\
\hline
 10&$10^{4.5}$&$10^{9}$&$10^{16.9}$&41.1&$5.4 \times10^{39}$\\
\hline
10&$10^5$&$10^{8.9}$&$10^{16.74}$&41.4&$1.1 \times10^{39}$\\
\hline
10&$10^{7}$&$10^{9}$&$10^{16.4}$&41.7&$8.4\times10^{37}$\\
 \hline
10&$10^{10.9}$&$10^{9.7}$&$10^{15.63}$&41.9&$3.2\times10^{34}$\\
 \hline
5&$10^{7.8}$&$10^{8.8}$&$10^{16.4}$&41.5&$9 \times10^{37}$\\
\hline
\end{tabular}
\caption{Allowed values of mass scales as solutions of RGEs for gauge couplings for Model-I with fixed value of the LH
  triplet mass $M_{\Delta}=10^8$~ GeV,}  
  \label{tab:gauge-gut1}
\end{table}
The RG evolution of gauge couplings for the set of mass scales given in
eq.(\ref{mod1sol}) is presented in Fig.\ref{fig:unifg2213} showing clearly the unification of the 
four gauge couplings of the $G_{2213}$ intermediate gauge symmetry.
 \begin{figure}[htbp]
\begin{center}
\includegraphics[width=7cm,height=7cm]{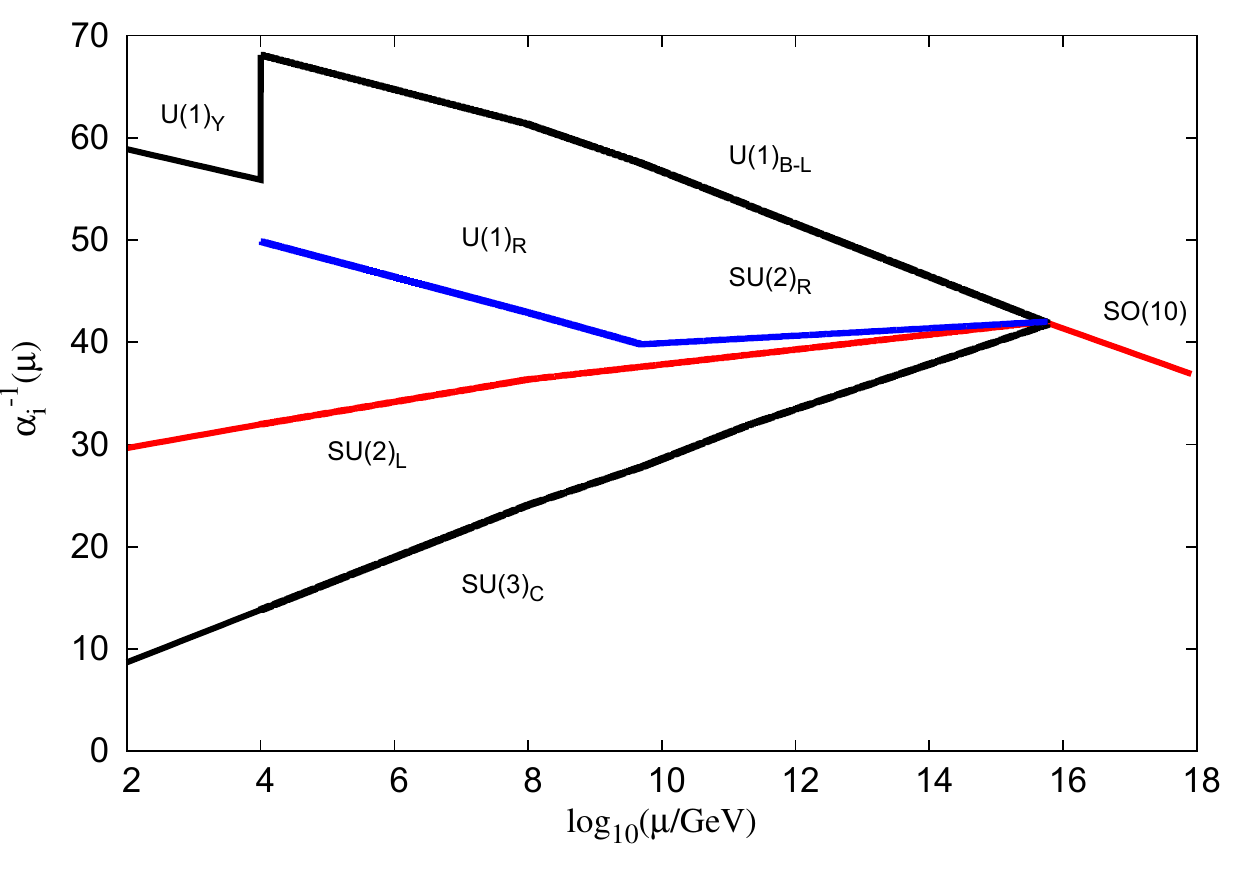}
\end{center} 
\caption{Two loop gauge coupling unification in the $SO(10)$ symmetry breaking chain with $M_U=10^{15.62}$ GeV and $M^+_R=10^{9.7}$, $M_{\Delta_L}=10^{8}$GeV with
 a low mass $Z^{\prime}$ boson at $M_R^0=10$ TeV for Model-I.}
\label{fig:unifg2213}
\end{figure}
 \par\noindent {\bf Model-II:}\\
In addition to the Higgs scalars of Table \ref{tab:particle},
this model has  the masses of left handed scalars $\chi_L(2,1,-1,1)$ and
 $\sigma_L(3,1,0,1)$ naturally at $\mu=M_R^+=M_P=$ the parity violation scale.}
 As shown in Table \ref{tab:gauge-gut2} for Model-II, the $G_{2213D}$
 symmetry is found to survive down to $M_{R^+}=M_P = 10^{8.2}$ GeV
 with $M_U= 6.5\times 10^{15}$ GeV when the color octet mass is at $M_C=10^8$ GeV. As discussed below, the proton lifetime in this case is closer to the current experimental limit.
\begin{table}[htbp]
\centering
\begin{tabular}{|l|l|l|c|c|c|}
\hline
$M_R^0$&$M_C$&$M^+_R$&$M_G$&$\alpha_G^{-1}$ & $\tau_p$\\
(TeV)&(GeV)&(GeV)&(GeV)&&(Yrs.)\\
\hline
 10&$10^{4.5}$&$10^{7.886}$&$10^{16.15}$&40.25&$4.3 \times10^{36}$\\
\hline
10&$10^{5.5}$&$10^{7.89}$&$10^{16.04}$&40.64&$1.6 \times10^{36}$\\
\hline
10&$10^8$&$10^{8.789}$&$10^{15.62}$&41.49&$3.9\times10^{34}$\\
\hline
10&$10^{8.5}$&$10^{8.8}$&$10^{15.5}$&41.69&$1.12\times10^{34}$\\
\hline
 5&$10^{5.8}$&$10^{7.2}$&$10^{15.83}$&41.15&$2.3 \times10^{35}$\\
\hline
\end{tabular}
\caption{Allowed mass scales as solutions of renormalisation group
  equations for Model-II as described in the text.}
  \label{tab:gauge-gut2}
\end{table}
\begin{figure}[htbp]
\begin{center} 
\includegraphics[width=7cm,height=7cm]{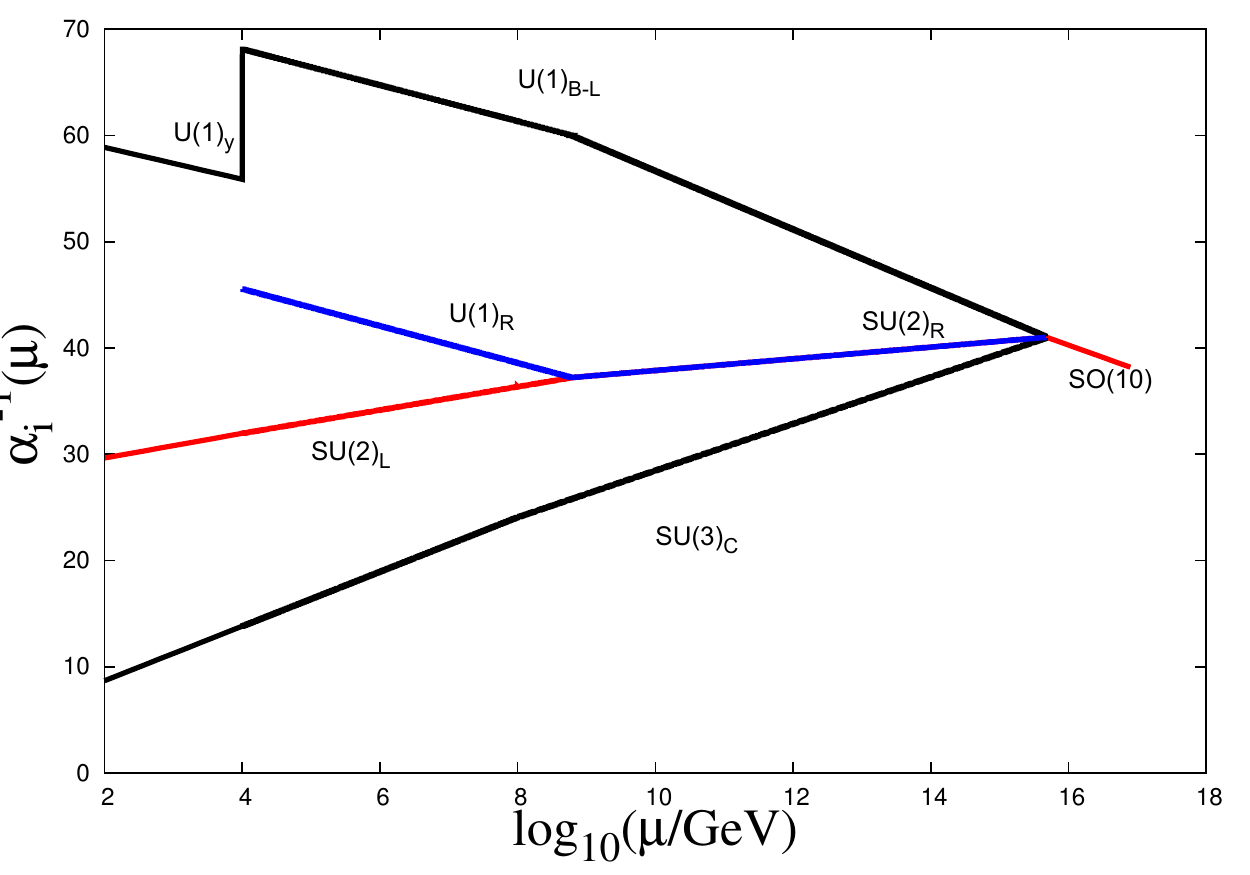}
\end{center}  
\caption{Two loop gauge coupling unification in the $SO(10)$ symmetry breaking 
chain with $M_U=10^{15.62}$ GeV and $M_{R^+}=10^{8.7}$ GeV with
 a low mass $Z^{\prime}$ boson at $M_R^0=10$ TeV for Model-II.}
\label{Fig1p}
\end{figure}
One example of RG evolution of gauge couplings is shown in 
Fig.\ref{Fig1p} for $M_{R^0}=10$ GeV, $M_{R^+}=10^{8.7}$ GeV, $M_C=10^8$ GeV,  and $M_U=6.5\times 10^{15}$ GeV. Clearly the figure shows precise unification of the three gauge couplings of the intermediate gauge symmetry $G_{2213P}$ at the GUT scale. For all other solutions given in Table-I, the RG evolutions and unification of gauge couplings are similar.
In both the models, with allowed values of $M_{R^+}\gg M_{R^0}= 5-10$
TeV, the numerical values of gauge couplings $g_{2L},g_{1R}$ and
$g_{B-L}$ predict \cite{langackerzp},
\be    
M_{Z^{\prime}} = (1.2 - 3.5) {\rm TeV}.\label{Mzprime} 
\ee
\subsection{Proton lifetime prediction}
\begin{figure}[htbp]
\begin{center}
\centering
\includegraphics[width=0.35\textwidth,height=0.35\textheight,angle=-90]{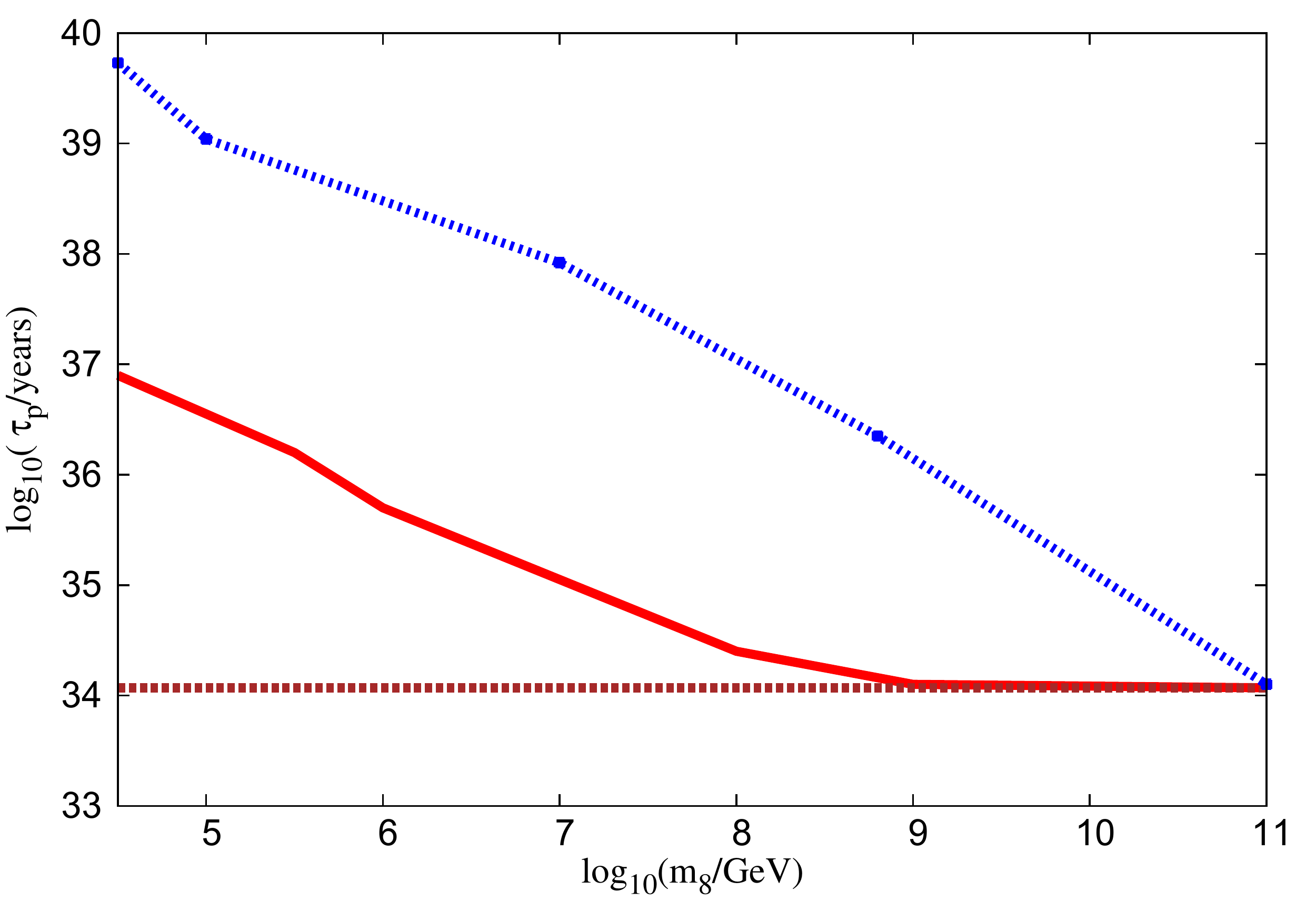}
\caption{Variation of proton lifetime as a function of  color octet
  mass $M_C$ for Model-I (upper curve) and Model-II (lower curve). The
horizontal line is the present experimental limit.}
\label{fig:taupvsm8}
\end{center}
 \end{figure}

In this section we discuss predictions on proton lifetimes in the two models and compare them with the current Super-Kamiokande limit and reachable limits by 
future experiments such as the Hyper-Kamiokande
\cite{babu11}. Currently, the Super-Kamiokande detector
has reached the search limit
\ba
{(\tau_p)}_{expt.}(p \to e^+\pi^0)&\ge & 1.4\times
  10^{34}~~{\rm yrs},\label{Superk}
\ea
The proposed $5.6$ Megaton years Cherenkov water detector at
Hyper-Kamiokande is expected to probe into lifetime \cite{babu11},
\ba
{(\tau_p)}_{Hyper-K.}(p \to e^+\pi^0) &\ge& 1.3\times10^{35}~~{\rm yrs},\label{Hyperk}
\ea
 The width of the proton decay for $p\to e^+\pi^0$ is expressed as \cite{bajc08}
 \bea   
\Gamma(p\rightarrow e^+\pi^0) 
&=&\left(\frac{m_p}{64\pi f_{\pi}^2}\right)\times
\left(\frac{{g_G}^4}{{M_U}^4}\right)\nonumber\\
&&|A_L|^2|\bar{\alpha_H}|^2(1+D+F)^2\times R.
\label{width}
\eea
where  $R=[(A_{SR}^2+A_{SL}^2) (1+ |{V_{ud}}|^2)^2]$ for $SO(10)$, $V_{ud}=0.974=$ the  
$(1,1)$ element of $V_{CKM}$ for quark mixings, $A_{SL}(A_{SR})$ is the short-distance 
renormalisation factor in the left (right) sectors and $A_L=1.25=$ long distance 
renormalization factor. $M_U=$ degenerate mass of $24$ superheavy gauge bosons in
$SO(10)$, $\bar\alpha_H =$ hadronic matrix element, $m_p = $ proton mass $=938.3$ MeV, 
$f_{\pi}=$ pion decay constant $=139$ MeV, and the chiral Lagrangian parameters are 
$D=0.81$ and $F=0.47$. With $\alpha_H= \bar{\alpha_H}(1+D+F)=0.012$ GeV$^3$ obtained
from lattice gauge theory computations, we get  $A_R \simeq A_LA_{SL}\simeq
A_LA_{SR}\simeq 2.726$ for both the  models. The expression for the inverse decay rates for 
the  models is expressed as, 
  \ba
  \tau_p=\Gamma^{-1}(p\rightarrow e^+\pi^0)  
 &=&\frac{64\pi f_{\pi}^2}{m_p}
 \left(\frac{{M_U}^4}{{g_G}^4}\right)\times\nonumber\\\frac{1}{|A_L|^2|\bar{\alpha_H}|^2(1+D+F)^2\times R}.
\label{width}
\ea
where the factor  $F_q=2(1+|V_{ud}|^2)^2\simeq 7.6$ for $SO(10)$. Now using  the given
values of the model parameters the predictions on proton lifetimes for 
both the models are given in Table \ref{tab:gauge-gut1} and Table
\ref{tab:gauge-gut2}. We find that for proton lifetime predictions
accessible to Hyper-Kamiokande detector, it is necessary
to have a intermediate value of the color octet mass $M_C \ge 10^{8.6}{\rm GeV}$ in Model-II
and $M_C \ge  10^{10.8}{\rm GeV}$  in Model-I.
The predicted proton lifetime as a function of the color octet mass is shown
in Fig. \ref{fig:taupvsm8} both for Model-I and for Model-II.
These analyses suggest that low color octet mass in the TeV scale and
observable proton lifetime within the Hyper-Kamiokande limit are mutually exclusive. If ~LHC discovers color octet within its achievable energy range, proton decay searches would need far bigger detector than the Hyper-K detector. On the other hand the absence of color octet at the LHC would still retain the possibility of observing proton decay within the Hyper-K limit.  
 \section{\small \bf TYPE-II SEESAW DOMINANCE}
\label{Sec.3}
In this section we discuss prospects of having a type-II seesaw
dominated neutrino mass formula in the two $SO(10)$ based models
discussed in Sec.\ref{Sec.2}.
\subsection{Derivation of type-II seesaw formula} 
\begin{figure}[htbp]
\begin{center}
\centering
\includegraphics[width=0.40\textwidth,height=0.25\textheight,angle=2]{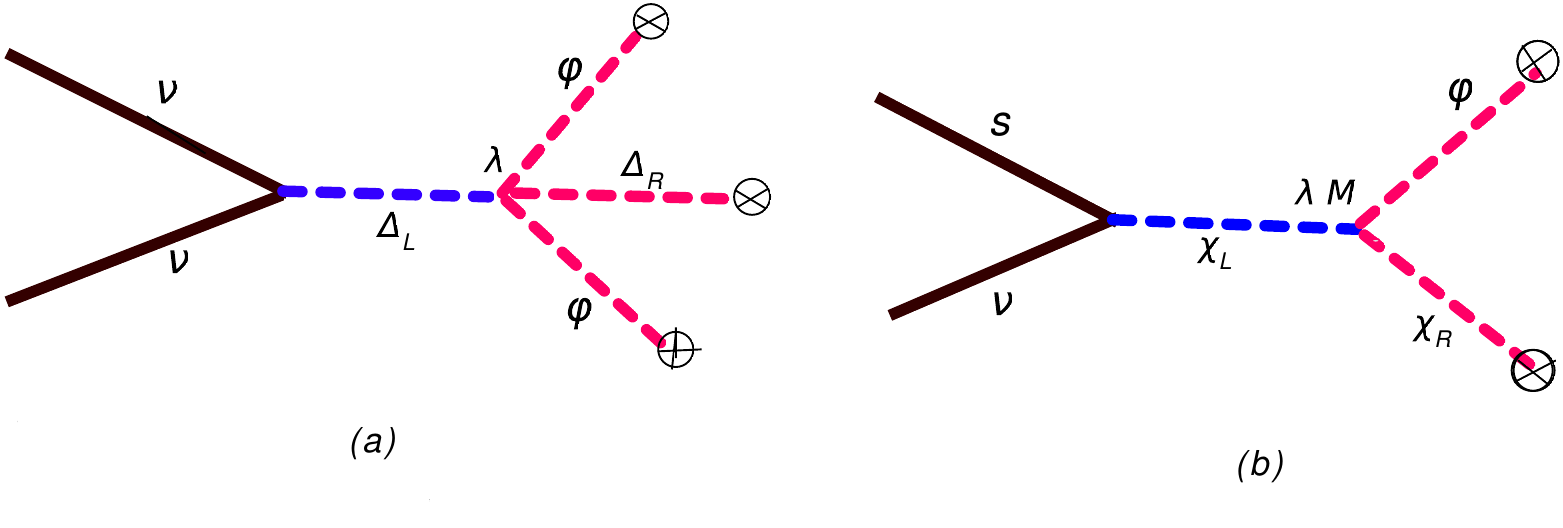}
\caption{Feynman diagrams for induced contributions to VEVs of the LH triplet (diagram (a)) and the LH doublet (diagram (b)) in Model-I and Model-II.}
\label{feynindvevs}
\end{center}
 \end{figure}

We have added to the usual spinorial representations ${16}_{F_i}(i=1,2,3)$ for fermion representations in $SO(10)$, one fermion singlet per generation $S_i(i=1,2,3)$. 
The $G_{2213}$ symmetric  Yukawa Lagrangian descending from $SO(10)$ symmetry can be written as
\begin{eqnarray}
\mathcal{L}_{\rm Yuk} &=& \sum_{i=1,2}Y_{i}^{\ell}\left (\overline{\psi}_L\, \psi_R\, \Phi_{i} \right)
+f(\, \psi^c_R\, \psi_R \Delta_R +\psi^c_L\, \psi_L \Delta_L)\nonumber\\
&+& y_{\chi}\,\left(\overline{\psi}_R\, S\, \chi_R +\overline{\psi}_L\, S\, \chi_L\right)
+\left(\text{h.c.}\right),\label{Yuk-Lag} 
\end{eqnarray}
where $\Phi_{1,2}\subset {10_{{H_1},{H_2}}}$ are two bidoublets, $(\Delta_L,\Delta_R)\subset {126}_F$ and $(\chi_L,\chi_R)\subset
{16}_H$. As discussed in Sec.2, the spontaneous breaking of $G_{2213}\longrightarrow G_{2113}$, takes place by the VEV of the RH triplet
$\sigma_R(1,3,0,1)\subset 45_H$ carrying $B-L=0$ which doesnot generate any fermion mass term. As we discuss below, when the Higgs scalar $\Phi_{i}$, $\Delta_R$ and $\chi_R$ acquire VEV's
spontaneous symmetry breakings leading to $G_{2113}\longrightarrow SM\longrightarrow U(1)_{em}\times SU(3)_C$ occur
 and generate $N-S$ mixing mass term $M= y_{\chi}\langle{\chi_R^0}\rangle$ by
the induced VEVs.In addition  $v_{\chi_L}=\langle{\chi_L^0}\rangle$ and $v_{L}=\langle{\Delta_L^0}\rangle$ are automatically generated even though the 
LH doublet $\chi_{L}$ and the RH triplet $\Delta_{L}$ are assigned vanishing VEVs directly.
 In models with inverse seesaw \cite{invseesaw} or extended seesaw
 \cite{ramlalmkp,app,pp,grimus} mechanisms,  a bare
 mass term of the singlet fermions $\mu_SS^TS$
occurs in the Lagrangian. Being unrestricted as a gauge singlet
mass term in the Lagrangian, determination of its value has been left
to phenomenological analyses in neutrino physics. Larger values of the
parameter near the GUT-Planck scale \cite{smirnov} or at the
intermediate scale \cite{barr} have been also exploited.
On the other hand, fits to the neutrino oscillation data through
inverse seesaw formula by a number of authors have shown to require 
much smaller values of $\mu_S$
\cite{app,pp,psbrnm,ramlalmkp,naturalness}. Even  phenomenological
implications of its vanishing value have been investigated recently in
the presence of other non-standard and non-vanishing fermion masses
\cite{hambyenu,psbpila} in the $9\times 9$ mass matrix. Very small
values 
of $\mu_S$ is justified on the basis of 't Hooft's naturalness
criteria representing a mild breaking of global lepton number symmetry
of the SM \cite{tHooft}. While we
consider the implication of this term later in this section, at
first we discuss the emerging neutrino mass matrix by neglecting it. 
In addition to the VEVs discussed in Sec.\ref{Sec.2} for gauge symmetry
breaking at different stages, we assign the VEV to the neutral
component of RH Higgs doublet of ${16}_H$ with
$<\chi_R(1,1/2,-1/2,1)>=V_{\chi}$ in order to generate $N-S$ mixing mass term
${\rm M} {\overline N}S$ between the RH neutrino and the sterile fermion where the
$3\times 3$ matrix ${\rm M}= y_{\chi}V_{\chi}$. We define the other  $3\times 3$
mass matrices $M_D=Y^{(1)}v_u$ and $M_N= fV_R$. We also include induced
small contributions to the vacuum expectation values of the LH Higgs triplet
 $v_L=<\Delta_L(3,0,-2,1)>$ and the LH Higgs doublet
$v_{\chi_L}=<\chi_L(2,0,-2,1)>$ leading to the possibilities
$\nu-S$ mixing with  $M_L=y_{\chi}v_{\chi_L}$ and the induced type-II seesaw contribution to LH
neutrino masses $m_{\nu}^{II}=fv_L$ given in eq.(\ref{indvevmod2}). The induced VEVs are shown in the
left and right panels of Fig.\ref{feynindvevs}. We have  also
derived them  by actual potential minimisation which agree  with the diagramatic contribution. Including the induced VEV contributions, the mass
term due to Yukawa Lagrangian can be written as 
\begin{eqnarray}
\mathcal{L}_{mass}&=& (M_D \overline{\nu}N +\frac{1}{2}M_NN^{T}N +M\overline{N}S \nonumber \\
& & +\left. M_L\overline{\nu}S + h.c\right)+m_{\nu}^{II}{\nu^T}\nu
\end{eqnarray}
In the $(\nu, S, N^C)$ basis   
the  generalised form of the $9\times 9$ neutral fermion mass matrix after electroweak symmetry breaking can be written as 
\ba
{\cal M} =  
\begin{pmatrix} m_{\nu}^{II} & M_L & M_D  \\
M_L^T & 0  & M^T\\ M_D^T & M & M_N \end{pmatrix}.  \label{fullnumatrix}
\ea
where $M_D=Y\langle \Phi\rangle$, $M_N=fv_R$, ${\rm
  M}=y_{\chi}\langle\chi_R^0\rangle$,$M_L=y_{\chi}\langle\chi_L^0\rangle$ and we have used $\mu_s=0$. 
In this model the symmetry breaking mechanism and the VEVs are such that $M_N > M \gg M_D$. The RH neutrino mass being the heaviest fermion mass scale in the Lagrangian, this fermion is at first integrated out
leading to the effective Lagrangian at lower scales \cite{kangkim,mkparc10,majee},
\begin{eqnarray}
- \mathcal{L}_{\rm eff} = \left(m_{\nu}^{II} + M_D \frac{1}{M_N} M^T_D\right)_{\alpha \beta}\, \nu^T_\alpha \nu_\beta +\nonumber\\
\left(M_{L} + M_D \frac{1}{M_N} M^T \right)_{\alpha m}\, \left(\overline{\nu_\alpha} S_m + \overline{S_m} \nu_\alpha \right)
\nonumber \\
 +\left(M \frac{1}{M_N} M^T\right)_{m n}\, S^T_m S_n, 
\end{eqnarray}
Whereas the heaviest RH neutrino mass matrix $M_N$ separates out trivially, the other two
  $3\times 3$ mass matrices $\mathcal{M}_\nu$, and  $m_S$
are extracted through various steps of block diagonalisation \cite{app}.The details of various steps are given in Appendix B and the results are
  \begin{eqnarray}
 \mathcal{M_\nu} =  m_{\nu}^{II} + \left(M_D M_N^{-1} M^T_D\right) -
(M_D M_N^{-1} M^T_D )\nonumber\\+ M_L(M^T M_N^{-1}M)^{-1} M_L^{T}\nonumber \\
 -M_L(M^T M_N^{-1}M)^{-1}(M^T M_N^{-1}M_D^T)\nonumber\\ -(M_D M_N^{-1} M) (M^T M_N^{-1}M)^{-1} M_L^{T},\nonumber\\
m_s =-MM_N^{-1}M^T+.... ,\nonumber\\
 m_N = M_N.\label{massmatrices} 
 \end{eqnarray} 
From the first of the above three equations, it is clear that the
type-I seesaw term cancels out \cite{kangkim,mkparc10,majee} with another of opposite sign
resulting from block diagonalisation. Then the generalised form of the  light neutrino mass matrix turns out to be
 \ba
{\cal M}_{\nu} &=&~ fv_L+M_LM^{-1}M_N(M^T)^{-1}M_L^T \nonumber\\
&&-[M_LM_D^TM^{-1} +M_L^TM_D(M^T)^{-1}]. 
 \label{inv}
\ea
 With $M_L=y_{\chi}v_{\chi_L}$ that induces $\nu-S$ mixing, the second
 term in this equation is double seesaw formula and the third term is
the linear seesaw formula which 
are similar to those derived earlier \cite{barr}. 
From the Feynman diagrams, the analytic expressions for the induced VEVs are
\ba 
v_L&\sim& \frac{V_R}{M_{\Delta_L}^2}\left(\lambda_1v_1^2+\lambda_2v_2^2\right),\nonumber\\  \label{vl}
\ea
\ba
v_{\chi_L}&\sim&\frac{V_{\chi}}{M_{\chi_L}^2}\left(\lambda^{\prime}_1M^{\prime}_1v_1+\lambda^{\prime}_2M^{\prime}_2v_2\right),\nonumber\\
&=&C_{\chi}\frac{V_{\chi}M_{R^+}v_{wk}}{M_{\chi_L}^2},\label{indvevs}
\ea
where $v_{wk}\sim 100$ GeV, and
\ba
C_{\chi}&=&{\left(\lambda^{\prime}_1M^{\prime}_1v_1+\lambda^{\prime}_2M^{\prime}_2v_2\right)\over (M_{R^+}v_{wk})}.\label{tunedfactor}
\ea
In eq.(\ref{indvevs}), $v_i(i=1,2)$ are the VEVs of two electroweak
doublets each originating from separate ${10}_H\subset SO(10)$ as
explained in the following section, and $M^{\prime}_1,M^{\prime}_2$
are Higgs trilinear coupling masses which are normally expected to be
of order $M_{R^+}$. In both the models $V_R=5-10$ TeV and $V_{\chi}\sim 300-1000$ GeV. Similar expressions as in eq.(\ref{indvevs}) are also obtained by minimisation of the scalar potential.

 \subsection{Suppression of linear seesaw and dominance of type-II seesaw}
 Now we discuss how linear seesaw term is suppressed without
 fine tuning of certain parameters in Model-I but with fine tunning 
of the same parameters in Model-II. The expression for neutrino mass
 is given in eq.(\ref{inv})
where the first, second, and the third terms are type-II seesaw,
double seesaw, and linear seesaw formulas for the light neutrino
masses. Out of these, for all parameters allowed in both the models
(Model-I and Model-II), the double seesaw term will be found to
be far more suppressed compared to the other two terms. Therefore we
now discuss how the linear seesaw term is suppressed compared to
the type-II seesaw term allowing the dominance of the latter.     
In Model-I, gauge coupling unification has been achieved such that $M_P=M_{\chi_L} \sim M_U\ge 10^{15.6}$ GeV, $M_{\Delta_L}=10^8$ GeV where $M^{\prime}_1\sim M^{\prime}_2\sim M_{R^+} \sim 10^9$ GeV. Using these masses in eq.(\ref{inv}), we find that even with $C_{\chi}\sim 0.1-1.0$
\ba
v_{\chi_L}&\sim& 10^{-18} ~{\rm eV}-10^{-17}~{\rm eV},\nonumber\\
v_L&\simeq& 0.1~{\rm eV} -  0.5~{\rm eV},\nonumber\\
\ea
Such induced VEVs in the Model-I suppress the second and the third terms in eq.(\ref{inv}) making the model  quite suitable for type-II seesaw dominance
although the  Model-II needs fine tuning in the induced contributions
to the level of $C_{\chi} \le 10^{-5}$ as discussed below.\\ 
 In Model-II, $M_{\Delta_L}\sim M_{\chi_L}\sim M_P\sim 10^9$ GeV
 , and without any fine tuning of the parameters in eq.(\ref{vl}), we obtain $v_L \sim 10^{-10}$~ GeV. From eq.(\ref{indvevs}) we get  $v_{\chi_L}\sim C_{\chi}\times 10^{-6}~{\rm {GeV}}\sim 10^{-7} {\rm {GeV}}$ for $C_{\chi}\sim 0.1$. With ${(M_D)}_{(3,3)}\leq 100$ GeV and ${M_D\over M} \simeq 0.1-1$, the most dominant third term in eq.(\ref{inv}) gives 
$M_{\nu}\geq 10^{-8}~{\rm GeV}$. This shows that fine tuning is
 needed in the parameters occuring to reduce  $C_{\chi}\le 10^{-5}$ to
 suppress linear seesaw and permit type-II seesaw dominance in
 Model-II whereas the type-II seesaw dominance is achieved in Model-I
 with $C_{\chi}\simeq 0.1-1.0$ without requiring any such fine tuning.
In what follows we will utilise the type-II seesaw dominated neutrino
mass formula to study neutrino physics\footnote{Following the similar block diagonalisation procedure given
  in Appendix B, but  in the
presence of  $\mu_SS^TS$ in the Yukawa Lagrangian with mass
ordering $M_N > M >> M_D, \mu_S$ results in the appearance of the
inverse seesaw part of the full neutrino mass matrix, \\ 
$ {\mathcal M}^{\prime}_{\nu} =~ fv_L+(\frac{M_D}{M})\mu_S{(\frac{M_D}{M})}^T.$\\
Although we plan to investigate the implications of this formula in a
future work, for the present purpose we assume $\mu_S\simeq 0$ 
such that type-II seesaw dominance prevails.},
neutrino-less double beta decay, and lepton flavor violations in the
context of Model-I although they are similar in Model-II subject to
the fine tuning constraint on $C_{\chi}$. Thus the light neutrino mass
is dominated by the type-II seesaw term 
\bea
{\mathcal M}_{\nu} &\simeq&~ fv_L.\label{indvevmod2}  
\eea
\subsection{Right-handed neutrino mass prediction} 
Global fits to the experimental data \cite{globalnu} on neutrino oscillations have determined the mass squared differences and mixing angles at $3\sigma$ level  
\begin{eqnarray}
&&\sin^2\theta_{12}=0.320,\,\,\,\, \sin^2\theta_{23}=0.427,\nonumber \\
&&\sin^2\theta_{13}=0.0246,\,\, \delta_{CP}=0.8 \pi,\nonumber \\
&&\Delta m_{\rm sol}^2=7.58\times 10^{-5}{\rm eV}^2, \nonumber\\
&&|\Delta m_{\rm atm}|^2=2.35\times 10^{-3}{\rm eV}^2.\label{oscdata}
\end{eqnarray}
For normally hierarchical (NH), inverted hierarchical (IH), and quasi-degenerate (QD) patterns, the experimental values of mass squared differences can be fitted by the following values of light neutrino masses
\ba
{\hat m}_{\nu}&=& (0.00127, 0.008838 ,0.04978) ~{\rm eV}\,\, (\rm {NH})\nonumber\\   
&=& ( 0.04901,0.04978,0.00127)~{\rm eV}\,\, (\rm {IH})\nonumber\\
&=& (  0.2056,0.2058,0.2) ~{\rm eV}\,\, (\rm {QD})\label{lnmasses}
\ea
We use the diagonalising Pontecorvo-Maki-Nakagawa-Sakata (PMNS) matrix.
The $U_{PMNS}$ matrix is give by
\bea
 \begin{pmatrix}
 c_{13}c_{12}&c_{13}s_{12}&s_{13}e^{-i\delta}\\
-c_{23}s_{12}-c_{12}s_{13}s_{23}e^{i\delta}&c_{12}c_{23}-s_{12}s_{13}s_{23}e^{i\delta}&s_{23}c_{13}\\
s_{12}s_{23}-c_{12}c_{23}s_{13}e^{i\delta}&-c_{12}s_{23}-s_{12}s_{13}c_{23}e^{i\delta}&c_{13}c_{23}\\
\end{pmatrix},\label{PMNS} 
\eea
and  determine it using mixing angle and the leptonic Dirac phase from eq. (\ref{oscdata})
\bea
U_{\rm {PMNS}}=\begin{pmatrix} 0.814&0.55&-0.12-0.09i\\
-0.35-0.049i&0.67-0.034i&0.645\\
0.448-0.057i&-0.48-0.039i&0.74\end{pmatrix}.\label{PMNSdata} 
\eea
Now inverting the relation ${\hat m}_\nu=U_{PMNS}^\dagger {\mathcal M}_\nu U_{PMNS}^*$
where ${\hat m}_\nu$ is the diagonalised neutrino mass matrix, we determine ${\mathcal M}_{\nu}$ for three different cases and further determine the corresponding values of the $f$ matrix using $f={\mathcal M}_\nu/v_L$
where we use the predicted value of $v_L=0.1$ eV. Noting that\\
$M_N=fV_R={{\mathcal M}_{\nu}V_R}/ v_L$, we have also derived eigen
values of the RH neutrino mass matrix  ${\hat M}_{N_i}$ as the
positive square root of the $i^{th}$ eigen value of the Hermitian matrix
$M_N^{\dag}M_N$.\\
 \noindent{\bf NH}\\
\ba
f =
\begin{pmatrix} 0.117+0.022i & -0.124-0.003i  &   0.144+0.025i\\
-0.124-0.003i  & 0.158-0.014i  & -0.141+0.017i\\ 0.144+0.025i &-0.141+0.017i& 0.313-0.00029i \end{pmatrix}\label{fNH}
\ea
\ba
 |{\hat M}_N| ={\rm diag}( 160, 894, 4870) ~{\rm GeV}.\,\ \label{MNNH}
\ea
\vspace{0.2cm}
\noindent{\bf IH}\\
\ba
f =  
\begin{pmatrix} 0.390-0.017i & 0.099+0.01i  &  -0.16+0.05i\\
0.099+0.01i  & 0.379+0.02i  & 0.176+0.036i\\-0.16+0.05i &0.176+0.036i& 0.21-0.011i \end{pmatrix}  \label{fIH}
\ea
\ba
 |{\hat M}_N| ={\rm diag}( 4880, 4910, 131) ~{\rm GeV}.\,\ \label{MNIH}
\ea
\noindent{\bf QD}\\
\ba
 f =
\begin{pmatrix} 2.02+0.02i & 0.0011+0.02i  &  -0.019+0.3i\\
0.0011+0.02i  & 2.034+0.017i  & 0.021+0.21i\\-0.019+0.3i &0.021+0.21i& 1.99-0.04i \end{pmatrix} \label{fQD}
\ea
For $v_L=0.1$ eV, we have 
\ba
 |{\hat M}_N| ={\rm diag}( 21.46, 20.34, 18.87) ~{\rm TeV}\ \label{MNQDp1}
\ea
but for  $v_L=0.5$ eV, we obtain
\ba
 |{\hat M}_N| ={\rm diag}( 4.3, 4.08, 3.77) ~{\rm TeV}.\ \label{MNQDp5}
\ea
These RH neutrino masses predicted with $v_L=0.1$ eV for NH and IH
cases and with $v_L=0.5$ eV for the QD case are clearly verifiable by the LHC.
\section{\small \bf THE DIRAC NEUTRINO MASS MATRIX }\label{Sec.4}
 The Dirac neutrino mass matrix which has its quark-lepton symmetric
 origin \cite{ps} plays a crucial role in the predictions of lepton
 flavor violations (LFVs) \cite {psbrnm,ramlalmkp}  as well as  lepton
 number violations (LNVs) as pointed out very recently \cite{app,pp}.      
The determination of the Dirac neutrino mass matrix $M_D(M_{R^0})$ at the TeV seesaw
scale is done which was discussed in \cite{ramlalmkp,dasparida}. \\
\subsection{Extrapolation to the GUT scale}
The RG extrapolated values at the GUT scale are, \\
\noindent{\large\bf{$\mu = {\rm M_{GUT}}$}:} 
\ba
m^0_e &=&0.00048~{\rm GeV}, m^0_{\mu}={\rm 0.0875}~{\rm GeV}, m^0_{\tau}=1.8739~
{\rm GeV},\nonumber\\
m^0_d&=&0.0027~{\rm GeV},m^0_s=0.0325~{\rm GeV}, m^0_b=1.3373~{\rm GeV},\nonumber\\ 
m^0_u&=&0.001~{\rm GeV}, m^0_c =0.229~{\rm GeV}, m^0_t = 78.74~{\rm GeV},            \label{eigenu}
\ea 
\vskip 0.2cm
The $V^0_{\rm CKM}$ matrix at the GUT scale is given by
\bea
V^0_{\rm CKM}= \footnotesize\begin{pmatrix} 0.97& 0.22& -0.0003-0.003i\\
-0.22-0.0001i& 0.97& 0.036\\
0.008-0.003i& -0.035+0.0008i& 0.99\end{pmatrix}.                 \label{vckmu} 
\eea 
For fitting the charged fermion masses at the GUT scale, in addition
to the two complex ${10}_{H_{1,2}}$ representations with their
respective Yukawa couplings $Y_{1,2}$, we also use the higher
dimensional operator \cite{psbrnm,ramlalmkp} 
\ba
{\bf \frac{\kappa_{ij}}{M_G^2}16_i16_j{10}_H{45}_H{45}_H},\label{dim6}
\ea
In the above equation the product of three Higgs scalars acts as an
effective ${126}_H^{\dagger}$ operator \cite{psbrnm}.
With $M_G\simeq M_{Pl}$ or $M\simeq  M_{string}$, this is suppressed by $(M_U/M_G)^2 \simeq 10^{-3}-10^{-5}$ for GUT-scale VEV of ${45}_H$.
Then the formulas for different charged fermion mass matrices are
\ba
M_u &=& {G}_u + {F}, ~~M_d ={G}_d + {F},\nonumber\\
M_e &=& {G}_d -3 {F},~~M_D ={G}_u -3 {F}.
\label{fmU}
\ea
Following the procedure given in \cite{ramlalmkp}, the Dirac neutrino mass matrix at the GUT scale is found to be
\bea
M_D(M_{R^0})=\footnotesize\begin{pmatrix}
0.014&0.04-0.01i&0.109-0.3i\\
0.04+0.01i&0.35&2.6+0.0007i\\
0.1+0.3i&2.6-0.0007i&79.20
\end{pmatrix}{GeV}.            \label{MDatMR0}
\eea 
\section{\small \bf LEPTON FLAVOR VIOLATION}\label{Sec.5}
In the present non-SUSY SO(10) models, even though neutrino masses are
governed by high scale type-II seesaw formula, the essential presence of
singlet fermions that implement the type-II seesaw dominance by
cancelling out the type-I seesaw contribution give rise to
experimentally observable LFV decay branching ratios through their
loop mediation. The heavier RH neutrinos in this model being in the
range of $\sim 1-10$ TeV mass range also contribute, but less
significantly than the singlet fermions.
The charged current weak interaction Lagrangian in this model can be
written in the generalised form,

\subsection{Estimation of non-unitarity matrix}
Using flavor basis, the general form of charged current weak interaction
Lagrangian including both $V \pm A$ currents in the Model-I and Model-II is 
\begin{eqnarray}
\mathcal{L}_{\rm CC} &=& -\frac{1}{\sqrt{2}}\, \sum_{\alpha=e, \mu, \tau}
\bigg[g_{2L} \overline{\ell}_{\alpha \,L}\, \gamma_\mu {\nu}_{\alpha \,L}\, W^{\mu}_L 
      + g_{2R}\overline{\ell}_{\alpha \,R}\, \gamma_\mu {N}_{\alpha \,R}\,
      W^{\mu}_R \bigg]\nonumber\\ 
   &&+ \text{h.c.} 
\label{eqn:ccint-flavor}
\end{eqnarray}
In both the models, the  $W^{\pm}_R$ bosons and the doubly charged Higgs
scalars, both left-handed (LH) and right handed (RH), are  quite heavy with $M_{W_R}\sim M_{\Delta}\simeq
10^8-10^9$ GeV. These make negligible contributions arising out of the RH current effects and Higgs exchange
effects on LFV or LNV decay amplitudes.  
In the two models considered here, the flavor eigenstate of any LH neutrino $\nu$ can be
represented in terms of mass eigen states $\nu_i$, $S_i$, and $N_i$.
From details of model parametrisation discussed in Sec.3-Sec.5, we have  found the corresponding mixing matrices with active
neutrinos, ${\cal V}^{\nu N}=M_D/M_N\equiv X_N$, and ${\cal V}^{\nu
  S}=M_D/M\equiv X_S$ 
\ba 
\nu &=&{\cal N}\nu_i+{\cal V}^{\nu N}N_i+{\cal V}^{\nu S}S_i, \nonumber\\
{\cal N}&\simeq& [1-(\eta^N+\eta^S)]U_{\rm {\small PMNS}}, \nonumber\\
\eta^N&=&(X_N.X_N^{\dagger})/2, \nonumber\\
\eta^S&=&(X_S.X_S^{\dagger})/2.\label{nonU}
\ea
 These
mixings modify the standard weak interaction Lagrangian in
the LH sector by small amounts but they could be in the experimentally
accessible range \cite{almeida:2000}. In the LH sector the charged
current weak interaction Lagrangian is
\ba
{\cal L}_{\rm CC}&=&-\frac{g_{2L}}{\sqrt 2}W_{\mu}{\bar
  e}\gamma^{\mu}P_L\left({\cal N}\nu_i+{\cal V}^{\nu N}N_i+{\cal
  V}^{\nu S}S_i\right)\nonumber\\
&&+h.c.\label{ModCC}
\ea
The implications of these terms for LFV and LNV effects have been
discussed below.
From eq.(\ref{nonU}) it is clear that ${\cal N}$ is non-unitary.
 We assume
the $N-S$ mixing matrix ${\rm M}$ to be diagonal for the sake of simplicity and economy of parameters, 
\ba
M ={\rm diag}~(M_1,M_2,M_3),\label{Mdiag}
\ea
Noting that the non-unitarity deviation is
characterised by $\eta=\eta^S+\eta^N$ which in the limit $M_N >> M$ turn out to be
\ba
\eta &\simeq&\eta^S= \frac{1}{2}X_S.X_S^{\dag}=M_DM^{-2}M_D^{\dag},\nonumber\\
\eta_{\alpha\beta}&=&\frac{1}{2}\sum_{k=1,2,3}\frac{M_{D_{\alpha k}}M_{D_{\beta
    k}}^*}{M_{k}^2}.\label{etaele}
\ea
For the degenerate case, $M_i=M_{Deg}(i=1,2,3)$, gives,
\bea
&&\eta =\frac{1 {\rm GeV}^2}{M_{Deg}^2}\times \nonumber \\
&& \footnotesize{\begin{pmatrix}
0.0394&0.146-0.403i&4.17-11.99i\\ 
0.146+0.403i&3.602&105.8-0.002i\\ 
4.173+11.9i&105.805+0.002i&3139.8
\end{pmatrix}.}\label{etaeq}
\eea
For the general non-degenerate case of  ${\rm M}$, we saturate the upper bound  $|\eta_{\tau\tau}|<2.7\times 10^{-3}$ \cite{antbnd} to derive
\ba
\frac{1}{2}\left[\frac{0.1026}{M_1^2}+\frac{7.0756}{M_2^2}+\frac{6762.4}{M_3^2}\right]
 = 2.7\times 10^{-3},\label{costr}
\ea
By inspection, this equation gives the lower bounds
\ba
M_1 &>& 4.35~{\rm GeV},~M_2 > 36.2~{\rm GeV}, M_3 > 1120 ~{\rm
GeV},\label{M123bounds}
\ea
and for the degenerate case $M_{Deg}=1213$ GeV.  
 For the partially degenerate case of $M_1=M_2\neq M_3$, 
the solutions can be similarly derived as in ref\cite{ramlalmkp} and
one example is $M (100,100,1319.67)$~GeV .
\subsection{Branching ratio and CP Violation}
One of the most important outcome of non-unitarity effects is expected to manifest 
through ongoing experimental searches for LFV decays such as  $\tau\to
e\gamma$, $\tau\to \mu\gamma$, $\mu\to e\gamma$. In these models the
RH neutrinos and the singlet fermions contribute to the branching
ratios \cite{psbrnm,ramlalmkp,ilakovac}.  
Because of the condition $M_N>> M$, neglecting the RH neutrino
exchange contribution  compared to the sterile fermion singlet
contributions, our estimations for different cases of $M$ values are
presented below. These values are many orders larger
than the standard non-SUSY contributions and are accessible to ongoing
or planned searches \cite{lfvexpt}.
For the degenerate case
\ba
\Delta {\cj}^{12}_{e\mu}&=&-2.1\times 10^{-6},\nonumber\\
\Delta {\cj}^{23}_{e\mu}&=&-2.4\times 10^{-6},\nonumber\\
\Delta {\cj}^{23}_{\mu\tau}&=&1.4\times 10^{-4},\nonumber\\
\Delta {\cj}^{31}_{\mu\tau}&=&1.2\times 10^{-4},\nonumber\\\label{cpvsol1}
\ea
 we have the predicted values of the branching ratios
\ba
BR(\mu \to e\gamma)&=&6.43\times 10^{-17},\nonumber\\
BR(\tau \to e\gamma)&=&8.0\times 10^{-16},\nonumber\\
BR(\tau \to \mu\gamma)&=&2.41\times 10^{-12}.\label{lfvbr}
\ea
Because of the presence of non-unitarity effects in the present model
, the leptonic CP-violation turn out to be similar as in refs.\cite{ramlalmkp,antbnd,non-unit}. The moduli and phase of non-unitarity and CP-violating parameter
for the degenerate case of the present models are\\
\ba
|\eta_{e\mu}|&=&2.73\times 10^{-8},\nonumber\\
\delta_{e\mu}&=&1.920,\nonumber\\
|\eta_{e\tau}|&=&4.54\times 10^{-7},\nonumber\\
\delta_{e\tau}&=&1.78,\nonumber\\
|\eta_{\mu\tau}|&=&2.31\times 10^{-5},\nonumber\\
\delta_{\mu\tau}&=&2.39\times 10^{-7}.\label{tabcpv}
\ea
The estimations presented in eq.(\ref{tabcpv})  show that in a wider
range of the parameter space, the leptonic CP violation parameter could
be nearly two orders larger than the CKM-CP violation parameter for quarks.  
\section{\small\bf NEUTRINO-LESS DOUBLE BETA DECAY}
Even with the vanishing bare mass term $\mu_S=0$ in the Yukawa
Lagrangian of eq.(\ref{Yuk-Lag}), the singlet fermions $S_i (i=1,2,3)$  acquire  Majorana masses over a
wide range of values and, in the leading order, the
corresponding mass matrix given in eq.(\ref{massmatrices}) is $m_S=
-M\frac{1}{M_N}M^T$. As far as light
neutrino mass matrix is concerned, it is given by the type-II seesaw
formula of eq.(\ref{indvevmod2}) which is independent of the Majorana
mass matrix $m_S$ of singlet fermions. But the combined effect of
substantial mixing between the light neutrinos and the singlet or the RH
neutrinos, and also between the singlet neutrinos and the RH neutrinos
result in the new Majorana neutrino mass
insertion terms in the Feynman diagrams. Out of these the mass
insertion $m_S$ due to the singlet fermions 
in the Feynman diagram gives rise to new dominant contributions to the
amplitude and the effective mass parameter for $0\nu\beta\beta$ even  in
the $W_L-W_L$ channel. This may be contrasted with the conventional
type-II seesaw dominated non-SUSY $SO(10)$ models with only three
generations of standard fermions in ${\bf {16}_i}(i=1,2,3)$ where
there is no such contributions to $0\nu\beta\beta$ decay. 
The generalised form of charged current interaction Lagrangian for
leptons in this model including both $V \pm A$ currents has
been given in eq.(\ref{eqn:ccint-flavor}).

\begin{figure}[htb!]
\includegraphics[width=5cm,height=4cm]{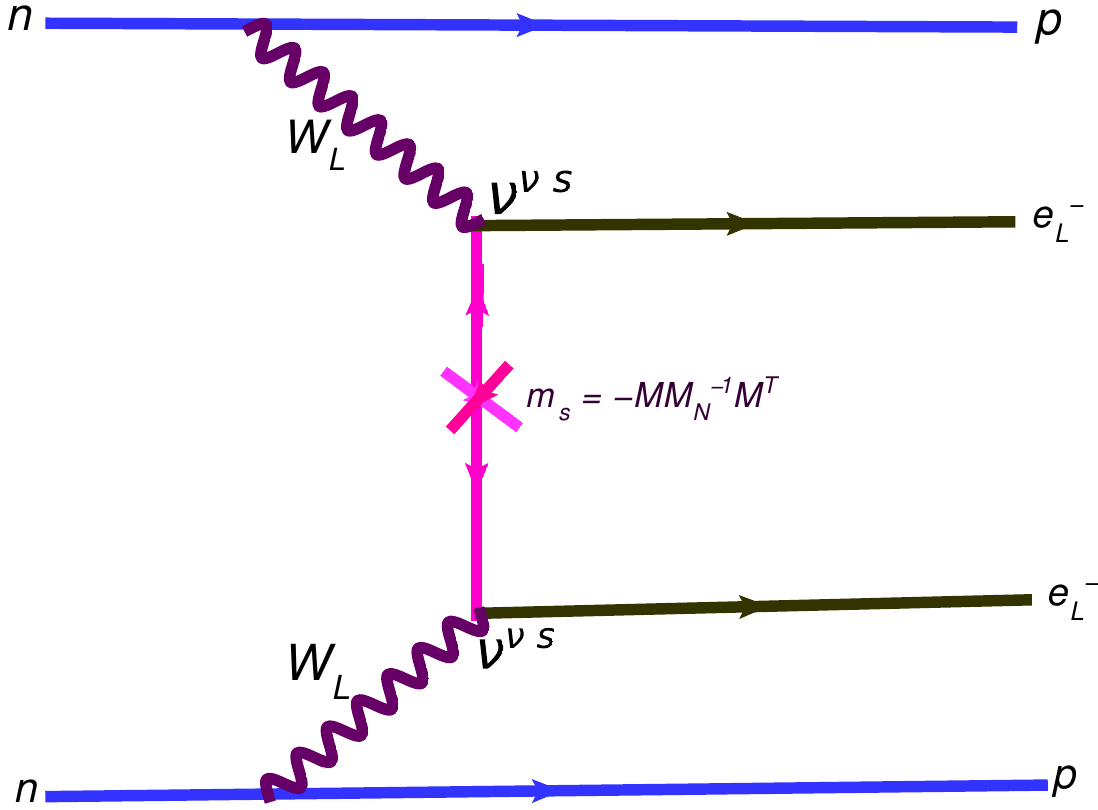}
\caption{Feynman diagrams for neutrinoless double beta decay  contribution 
with virtual Majorana neutrinos $\hat{\nu}_i$, and  $\hat{S}_i$
in the $W_L-W_L$-channel. For the RH neutrino exchange the Feynman diagram is same as in the right-panel but with the replacements ${\mathcal V}^{\nu S} \to {\mathcal V}^{\nu N}$, $S_i\to N_i$}
\label{t2beta}
\end{figure}

As stated above, in the Model-I and Model-II, the  $W^{\pm}_R$ bosons and the doubly charged Higgs
scalars, both left-handed and the right handed,  are  quite heavy with $M_{W_R}\sim M_{\Delta}\simeq
10^8-10^9$ GeV. These make negligible contributions due to the RH current effects and Higgs exchange
effects for the $0\nu\beta\beta$ decay amplitude.  
The most popular standard and conventional contribution in the $W^-_L - W^-_L$
channel is due to  light neutrino exchanges.
But one major new point in this work 
is that even in the $W^-_L - W^-_L$ channel, the singlet fermion exchange 
allowed within the type-II seesaw dominance mechanism, can yield much more dominant 
contribution to $0\nu \beta \beta$ decay rate. For 
the exchange of  singlet fermions
($\hat{S}_j$), the Feynman diagram is shown in the 
Fig.\ref
{t2beta}. For the exchange of heavier RH  Majorana neutrinos
($\hat{N}_k$), the diagram is the same as the right-panel of this figure
but with the replacement of the  mixing matrix and masses  by
${\mathcal V}^{\nu S} \to {\mathcal V}^{\nu N}$ and $m_{S_i}\to M_{N_i}$. 
The heavier RH neutrino exchange contributions are found to be
negligible compared to the singlet fermion exchange contributions.
In the mass basis, the
contributions to the decay amplitudes by $\nu$ and $S$ exchanges are estimated as 
\begin{eqnarray}
\label{eq:amp_ll} 
& &\mathcal{A}^{LL}_{\nu} \propto \frac{1}{M^4_{W_L}} \sum_{i=1,2,3} \frac{\left(\mathcal{V}^{\nu \nu}_{e\,i}\right)^2\, m_{\nu_i}}{p^2} 
\end{eqnarray}
\bea
& &\mathcal{A}^{LL}_{S} \propto \frac{1}{M^4_{W_L}} \sum_{j=1,2,3} \frac{\left(\mathcal{V}^{\nu S}_{e\,j}\right)^2}{m_{S_j}}
\label{samp}
\eea
\begin{eqnarray}
& &\mathcal{A}^{LL}_{N} \propto \frac{1}{M^4_{W_L}} \sum_{j=1,2,3} \frac{\left(\mathcal{V}^{\nu N}_{e\,j}\right)^2}{m_{N_j}},\label{sampt}  
\end{eqnarray}
where $|p| \simeq 190~\mbox{MeV}$ represents the
magnitude of neutrino virtuality momentum \cite{nuvirt-moh,Doi}. Using
uncertainities in the nuclear matrix elements \cite{Vergados,Doi} we
have found it to have values in the range $|p|= 120 {\rm MeV} - 200
{\rm MeV}$.
 In order to understand physically how the singlet fermion  Majorana mass insertion terms as
a new source of lepton number violation contributes  
to $0\nu\beta\beta$ process, we draw the Feynman diagram Fig.\ref{t2beta}. 
with mass insertion.  
In this model, the Majorana mass matrix for the singlet fermion after block diagonalisation is  
 $m_S=-M M_N^{-1}M^T$. Then exchanges of such  singlets generate
 dominant contribution through their mixings to active neutrinos and
 this mixing is proportional to the Dirac neutrino mass $M_D$ derived
 in \ref{Sec.4}. It is clear from  
 Fig. \ref{t2beta}
 that the the singlet fermion exchange amplitudes assume 
 the same form as 
 in eq.(\ref{samp}).
\section{\small\bf{EFFECTIVE  MASS PARAMETER AND HALF LIFE}}
Adding together the $0\nu\beta\beta$ decay amplitudes arising out
of light neutrino exchanges, singlet fermion exchanges, and the heavy
RH neutrino exchanges in the $W_L-W_L$ channel from eq.(\ref{samp}), and using suitable normalisations
\cite{Vergados,Doi}, we express the inverse half life   
\begin{eqnarray}
  \left[T_{1/2}^{0\nu}\right]^{-1} &\simeq &
  G^{0\nu}_{01}|\frac{{\cal M}^{0\nu}_\nu}{m_e}|^2|({\large\bf
    m}^{ee}_{\nu} +{\large\bf m}^{ee}_{S}+{\large\bf m}^{ee}_{N})|^2,\nonumber\\
&=& K_{0\nu}|({\large\bf
    m}^{ee}_{\nu} +{\large\bf m}^{ee}_{S}+{\large\bf m}^{ee}_{N})|^2,\nonumber\\
&=& K_{0\nu}|{\large\bf
    m}_{\rm eff} |^2
  \label{eq:invhalflife_simp}
\end{eqnarray}
In the above equation  $G^{0\nu}_{01}= 0.686\times 10^{-14} {\rm yrs}^{-1}$, ${\cal
  M}^{0\nu}_{\nu} = 2.58-6.64$, $K_{0\nu}= 1.57\times 10^{-25} {\rm yrs}^{-1}
{\rm eV}^{-2}$, and the three effective mass parameters for
light neutrino, singlet fermion, and heavy RH neutrino exchanges  are 
\begin{eqnarray}
{\large \bf  m}^{\rm ee}_{\nu} =\sum^{}_{i} \left(\mathcal{V}^{\nu \nu}_{e\,i}\right)^2\, {m_{\nu_i}}
\label{effmassparanus} 
\end{eqnarray}
\bea
{\large \bf  m}^{\rm ee}_{S} = \sum^{}_{i} \left(\mathcal{V}^{\nu
  S}_{e\,i}\right)^2\, \frac{|p|^2}{m_{S_i}} 
  \label{effmassparanus2} 
\eea
\begin{eqnarray}
{\large \bf  m}^{\rm ee}_{N} = \sum^{}_{i} \left(\mathcal{V}^{\nu
  N}_{e\,i}\right)^2\, \frac{|p|^2}{m_{N_i}},
\label{effmassparanus3} 
\end{eqnarray}
with 
\begin{eqnarray}
&&{\large\bf m}_{\rm eff}={\large\bf m}^{ee}_{\nu} +{\large\bf
    m}^{ee}_{S}+{\large\bf m}^{ee}_{N}.\label{sumeff}
\end{eqnarray}
Here $m_{S_i}$ is the eigen value of the $S-$ fermion mass matrix
$m_S$, and the magnitude of neutrino virtuality momentum
$|p|= 120$ MeV$-200$ MeV.
As the predicted values of the RH neutrino masses carried out in Sec.3
have been found to be large which make their contribution to the
$0\nu\beta\beta$ decay amplitude negligible,  we retain only
contributions due to light neutrino and singlet fermion exchanges.  
The estimated values of the effective mass parameters due to the $S-$ fermion
exchanges and light neutrino exchanges are shown separately 
in Fig. \ref{fig:effmassp}
where  the magnitudes of corresponding mass eigen values used have
been also indicated.
\subsection {Numerical estimations of effective mass parameters}
Using  the equations of normalized mass parameters \cite{app} , we estimate  numerically  the nearly standard contribution due to light neutrino exchanges and  the dominant non-standard contributions due to singlet fermion exchanges.
\par \noindent{\bf A.Nearly standard contribution}\\
In this model the new mixing matrix $\mathcal{N} \equiv \mathcal{V}^{\nu \nu}= \left(1- \eta \right) U_{\nu}$ contains additional non-unitarity effect due to non-vanishing $\eta$ \cite{app}
Using $M_{Deg}= 1213$~GeV in the degenerate case, we estimate
\be
{\mathcal{N}}_{ei} = ~(0.81437,~0.54858,~0.1267+0.0922i).\label{Nei}
\ee
Since all the $\eta-$ parameters are constrained by $|\eta_{\alpha\beta}| < 10^{-3}$, it is expected
that  $|{\mathcal{N}}_{ei}| \simeq |U_{ei}|$ for any other choice of $M$.
 In the leading approximation, by neglecting the $\eta_{\alpha i}$
 contributions , the effective mass parameter in the the $W_L-W_L$ channel with light
neutrino exchanges is expressed as
\bea 
m^{\rm ee}_{\nu} &=& \sum_i{\mathcal N}_{ei}^2{\hat  m}_i\nonumber\\
&&\simeq (c_{12}c_{13})^2{\hat m}_1e^{i\alpha_1}+(s_{12}c_{13})^2{\hat
  m}_2e^{i\alpha_2}\nonumber\\
&&+s_{13}^2e^{i\delta}{\hat m}_3,\label{std:meff}
\eea
where we have introduced two Majorana phases $\alpha_1$ and
$\alpha_2$. As discussed subsequently in this section, they play
crucial roles in preventing cancellation between two different
effective mass parameters.
Using  $\alpha_1=\alpha_2=0$ and the
experimental values of light neutrino masses and  the Dirac phase
$\delta=0.8\pi$ from eq.(\ref{oscdata}), the light neutrino exchanges have their
well known values,  
\bea
|m^{\rm ee}_{\nu}|
&&=\left\{
\baz 
0.0039\, \mbox{eV}
& \mbox{ NH,} \\[0.2cm]
0.04805\, \mbox{eV}
& \mbox{ IH,} \\[0.2cm]
0.23\, \mbox{eV} 
& \mbox{ QD.} 
\end{array}
\right. 
\eea
 \vspace{0.2 cm}
\par
\begin{figure}[htb!]
\includegraphics[width=4cm,height=6cm, angle=-90]{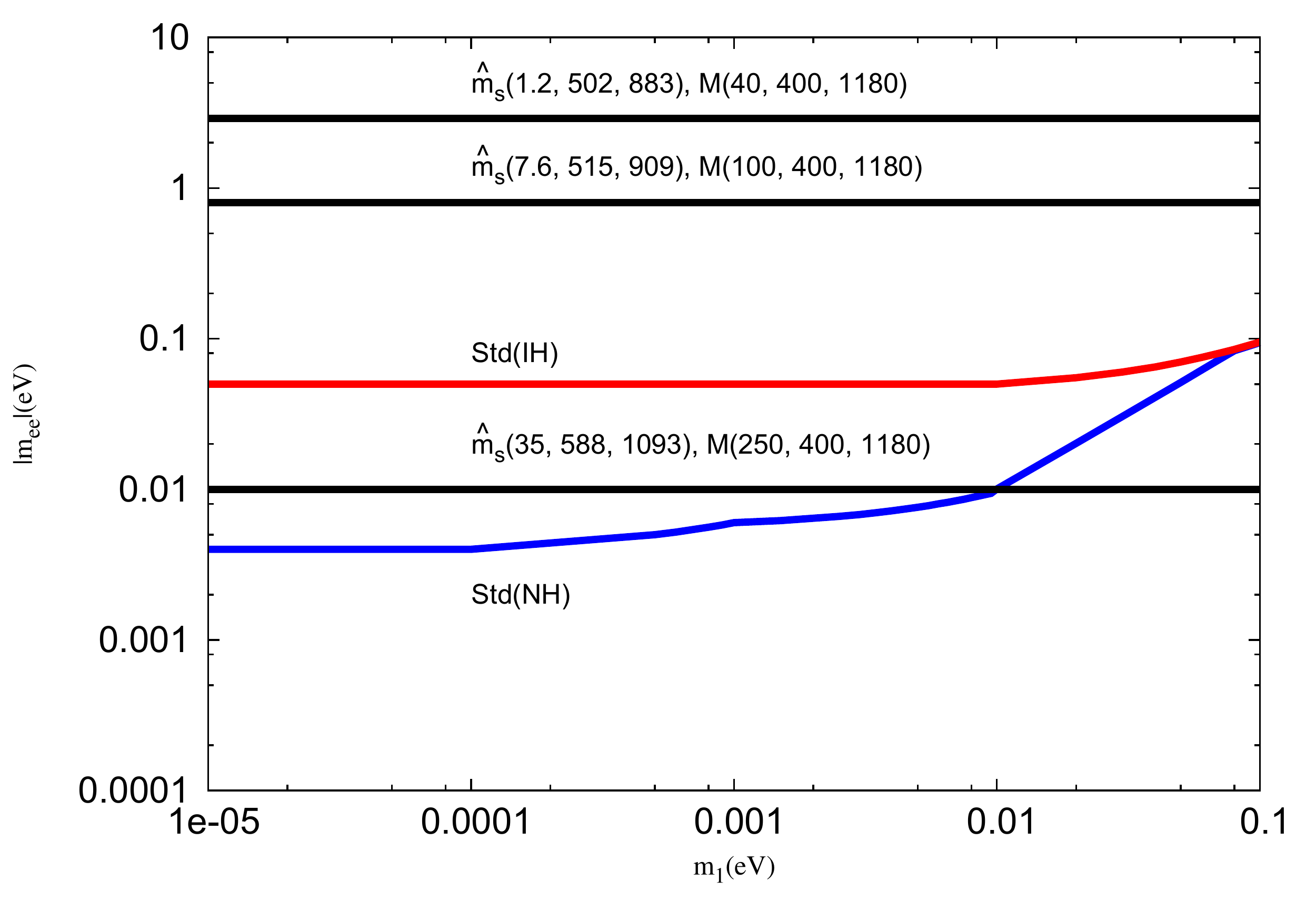}
\caption{Variation of the effective mass parameters with lightest
  LH neutrino mass. The dominant non-standard contributions due to
  fermion singlet contributions are shown by three horizontal lines with
  corresponding mass values in GeV units. The subdominant effective mass parameters
  due to NH and IH cases shown are similar to the standard values. }
   \label{fig:effmassp}
   \end{figure}

 \noindent{\bf B. Dominant non-standard contributions} \\ 
 The  $(ei)$ element of the $\nu-S$ mixing matrix is \cite{app}
\be
 {\mathcal{V}}^{\nu S}_{ei} = (\frac{M_D}{M})_{ei}.\label{vnus}   
\ee
where the Dirac neutrino mass matrix $M_D$ has been given in eq.(\ref{MDatMR0}), and
the diagonal elements are estimated using the non-unitarity equation
as discussed in the previous section.
We derive the relevant elements of the mixing matrix
$\mathcal{V}^{\nu S}$ using the structures of the
the Dirac neutrino mass matrix $M_D$ given in eq.(\ref{MDatMR0}) and
values of the diagonal elements of $M=(M_1, M_2, M_3)$ satisfying the
non-unitarity constraint in eq.(\ref{costr}). The eigen values of the
$S-$ fermion mass matrix $m_S$ are estimated for different cases using the
structures of the  
 RH Majorana neutrino
mass matrices given in eq.(\ref{MNNH}),
eq.(\ref{MNIH}), and eq.(\ref{MNQDp1}) in the formula
$m_S=-M\frac{1}{M_N}M^T$.
It is clear that in the effective mass parameter the non-standard contribution due to sterile fermion exchange
has a sign opposite to that due to light neutrino exchange and also its
magnitude is inversely proportional to the sterile fermion mass eigen values.    
In the NH case the estimated effective mass parameters are shown in
Fig.\ref{fig:effmassp} where the values of diagonal elements of $M$ and
the eigen values of $m_s$ have been specified. For comparison the
effective mass parameters in the standard case without singlet
fermions have been also given. It is clear that for allowed masses of
the model, the non-standard
contributions to effective mass parameters can be much more dominant
compared to the standard values irrespective of the mass patterns of light
neutrino masses:NH, IH, or QD.   
\subsection{Cancellation between effective mass parameters}  
 When plotted
as a function of singlet fermion mass eigen value $m_{S_1}$, the resultant
effective mass parameter shows cancellation for certain region of the
parameter space, the cancellation being prominent in the QD case.     
Like the light neutrino masses, the singlet fermion masses $m_{S_i}$
are also expected to have two Majorana phases. When all Majorana
phases are absent, both in the light active neutrino as well as in the singlet fermion sectors,
it is clear that in the sum of the two effective mass parameter there will
be cancellation between light active neutrino and the singlet fermion contributions because of
the inherent negative sign of the non-standard contribution. Our
estimations for NH, IH, and QD patterns of light neutrino mass
hierarchies are discussed separately.
 \par
\begin{figure}[htbp]
\includegraphics[width=6cm,height=4cm]{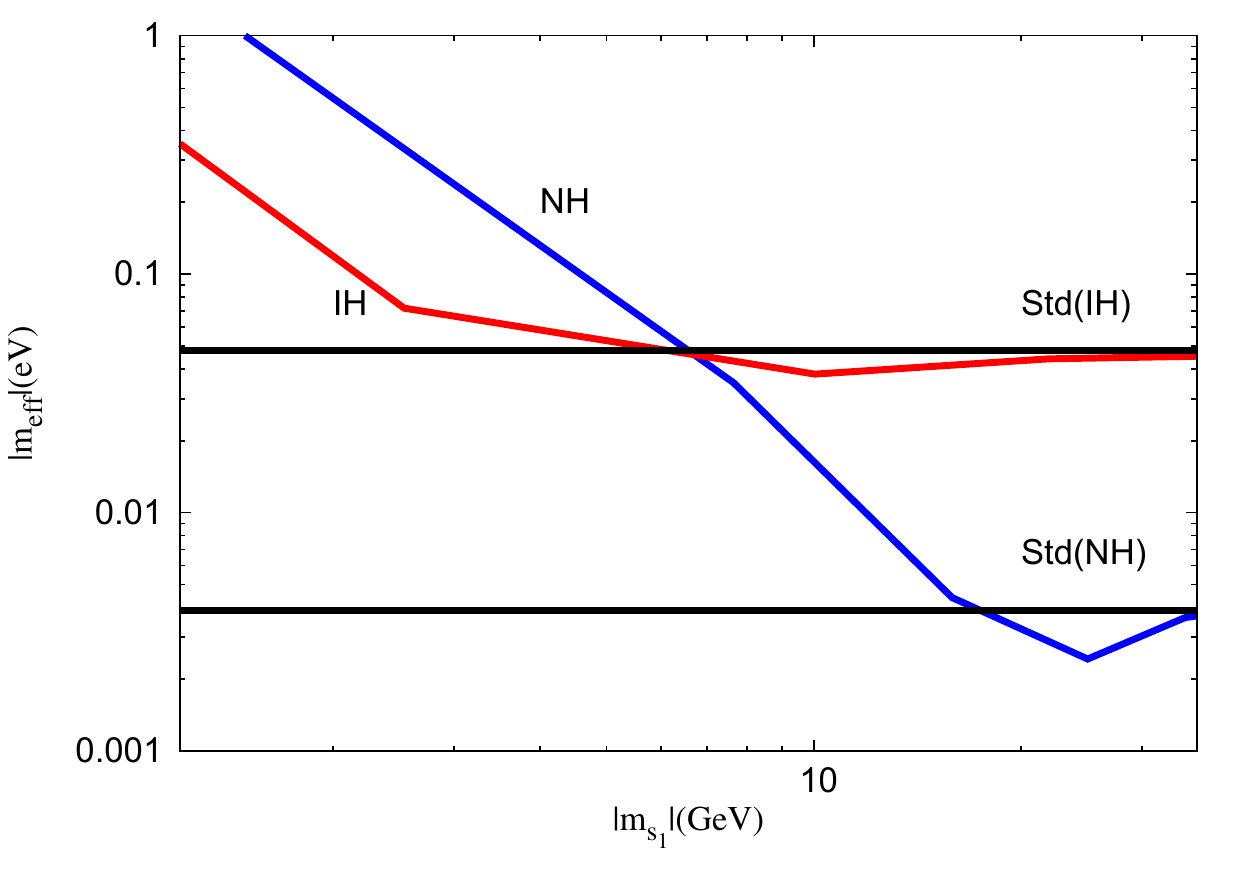}
\caption{Variation of effective mass parameter of $0\nu\beta\beta$ decay with the
  mass of the lightest singlet fermion for  $|p|=190$ {MeV} .}
\label{cmeffs1}
\end{figure}

\noindent{\bf A. Effective mass parameter for NH and IH active
  neutrino masses}\\
In Fig. \ref{cmeffs1}, we have shown the variation of the resultant effective mass parameter with $m_{S_1}$ for NH
and IH patterns of active light neutrino masses. It is clear
that for lower values of $m_{S_1}$, the singlet fermion exchange term
continues to dominate. For larger values of
$m_{S_1}$ the resultant effective mass parameter tends to be identical
to the light neutrino mass contribution due to the vanishing
non-standard contribution. We note that the values $|m_{eff}|=0.5-0.1$
eV can be easily realised for $|m_{S_1}|= 3-5$ GeV in the NH case but
for  $|m_{S_1}|= 1-2$ GeV in the IH case.  
\par
\begin{figure}[htbp]
\includegraphics[width=6cm,height=4cm]{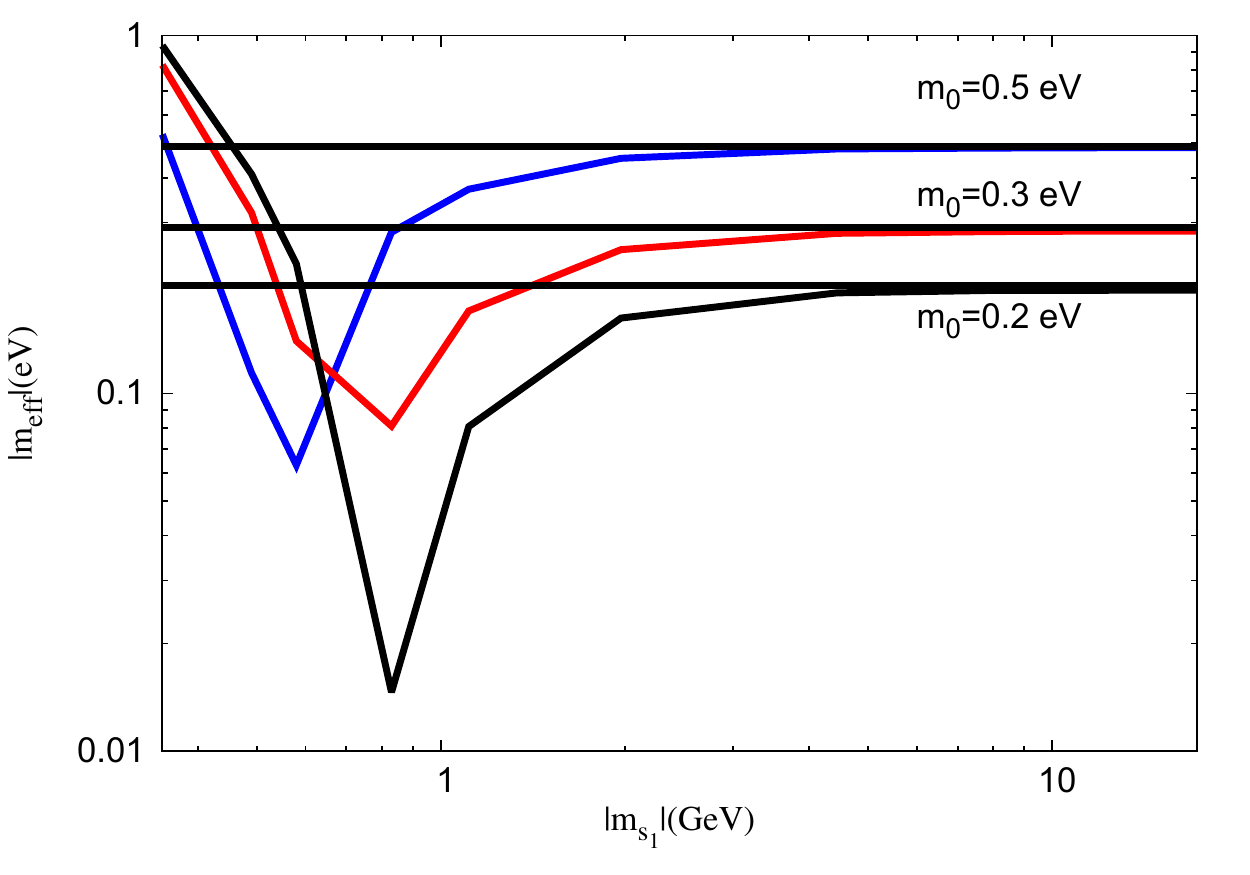}
\includegraphics[width=6cm,height=4cm]{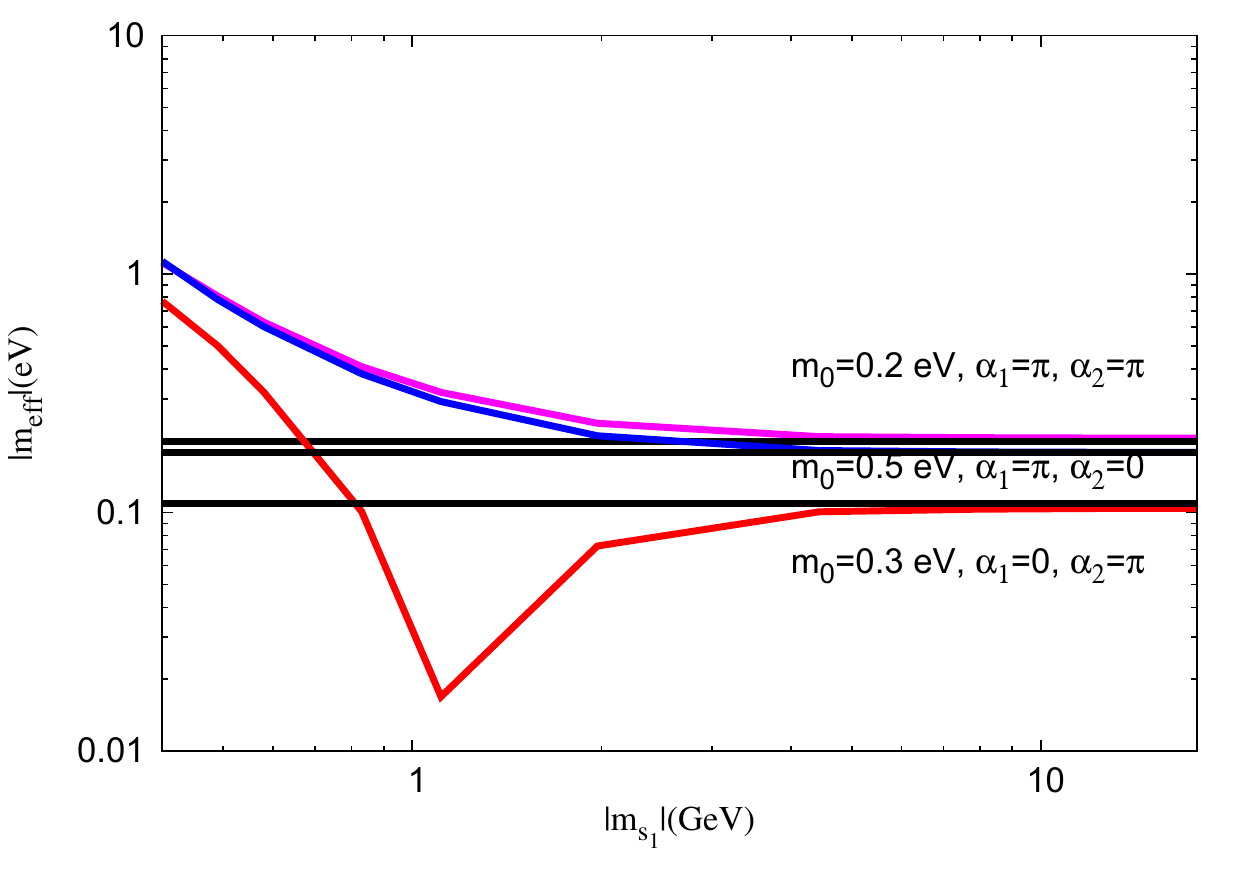}
\caption{Variation of effective mass of $0\nu\beta\beta$ decay with the
  mass of the lightest singlet fermion for QD light neutrinos with one Dirac phase (left), and with one Dirac phase and two Majorana phases (right)
 .}
\label{cmeffsqd}
\end{figure}

 \noindent{\bf B. Effective mass parameter for QD neutrinos}\\    
The variation of effective mass with 
 $m_{S_1}$ for the QD case with one experimentally determined Dirac
phase $\delta =0.8\pi$
 and
 assumed values of two unknown Majorana phases is given in Fig. \ref{cmeffsqd}.
The upper-panel of Fig. \ref{cmeffsqd} shows the variation 
with  $\alpha_1= \alpha_2 =0$ for different choices of the common
light neutrino mass $m_0=0.5$ eV, $0.3$ eV , and $0.2$ eV for the
upper, middle, and the lower curves, respectively, where cancellations are clearly
displayed in the regions of $m_{s_1}=0.4-1.5$ GeV.
 However, before such cancellation occurs, the dominance of
the singlet exchange contribution has been clearly shown to occur in
the regions of lower values of $m_{S_1}$. For larger values of
$m_{S_1}> 5 $ GeV, the singlet exchange contribution tends to be
negligible and the light QD neutrino contribution to $m_{eff}$ is recovered.
 In the lower panel of Fig. \ref{cmeffsqd}, the upper curve corresponds to  $\alpha_1=\pi$,$\alpha_2=\pi$ at $m_{0}=0.2 eV$.
 The middle line corresponds to $\alpha_1=\pi$,$\alpha_2=0$ at
 $m_{0}=0.5eV$ .The lower line corresponds to
 $\alpha_1=0$,$\alpha_2=\pi$ at $m_{0}=0.3 eV$.
We find that because of introduction of appropriate Majorana phases
the dips in two curves have disappeared. 
\subsection{Half-life as a function of singlet fermion masses}
In order to arrive at a plot of half-life against the lightest singlet
fermion mass in different cases, at first we estimate the mass eigen
values of the three singlet fermions for different allowed
combinations of the $N-S$ mixing matrix elements satisfying the
non-unitarity constraint of eq.(\ref{costr}) and by using the RH
neutrino mass matrices predicted for NH, IH, and QD cases from
eq.(\ref{MNNH}), eq.(\ref{MNIH}), eq.(\ref{MNQDp1}), and eq.(\ref{MNQDp5}).
 These solutions are shown in Table \ref{tableMms}.\\
 We then derive expressions for half-life taking into account  the
contributions of the two different amplitudes or effective mass
parameters arising out of the light neutrino and the singlet fermion
exchanges leading to
\be
\left[T_{1/2}^{0\nu}\right]=\frac{{m_{s1}^2}}{{K_{0\nu}}|p|^4{({M_{D}/M})_{e1}}^4}\bigg[|1+X+Y|\bigg]^{-2},\label{XYhalfbb}
\ee
where
\bea 
X&=&\frac{{{({M_{D}/M}})_{e2}}^2}{{{({M_{D}/M}})_{e1}}^2}\frac{m_{S_1}}{m_{s_2}}
+\frac{{{({M_{D}/M}})_{e3}}^2}{{{({M_{D}/M}})_{e1}}^2}\frac{m_{S_1}}{m_{S_3}},\\
Y&=&{\bf m}^{ee}_{\nu}\frac{m_{S_1}}{p^2 {(M_{D}/ M)}_{e1}^2}.
\label{XYdefinhalf} 
\eea
Here we have used the expression for ${\bf m}^{ee}_{\nu}$  given in
eq.(\ref{effmassparanus}). In eq.(\ref{XYhalfbb}), $Y=0$ gives
complete dominance of the singlet fermion exchange term. However this formula of half-life is completely different from the one
obtained using inverse seesaw dominance in SO(10) \cite{pas}. In the
present model in the leading order, the predicted half-life depends directly on the square of the
lightest singlet fermion mass and it is independent of the RH neutrino mass which is non-diagonal. But in \cite{pas}
, the half-life of neutrino less double beta decay is directly proportional to the fourth power of the lightest singlet fermion mass and square of the lightest
right handed neutrino mass leading into a different result.\\  
\begin{table*}
\begin{center}
\begin{tabular}{|c|c|}
\hline
$M$&$\hat{m_{s}}(NH)$\\
(GeV)&(GeV)\\ \hline
(40,400,1180)& (1.2,502,883)\\ \hline
(100,400,1180)& (7.65,515,909))\\ \hline
(150,400,1180)& (16,533,951)\\ \hline
(200,400,1180)& (25,558,1011)\\ \hline
(250,400,1180)& (35,588,1093)\\ \hline
(300,400,1180)& (43,622,1200)\\ \hline
(350,400,1180)& (50,659,1331)\\ \hline
\end{tabular}
\begin{tabular}{|c|c|}
\hline
$M$&$\hat{m_{s}}(IH)$\\
(GeV)&(GeV)\\ \hline
(40,450,1280)& (0.4,54.32,7702)\\ \hline
(60,450,1280)& (0.9,54.4,7705)\\ \hline
(70,450,1280)& (1.2,54.4,7706)\\ \hline
(100,450,1280)& (2.5,55,7715)\\ \hline
(300,450,1280)& (22,56,7831)\\ \hline
(400,450,1280)& (36.2,59,7933)\\ \hline
(450,450,1280)& (42,64,7996)\\ \hline
\end{tabular}
\begin{tabular}{|c|c|}
\hline
$M$&$\hat{m_{s}}(QD)$\\
(GeV)&(GeV)\\ \hline
(100,600,1500)& (0.5,17.7,109))\\ \hline
(130,600,1500)& (0.8,17.7,109)\\ \hline
(200,600,1500)& (1.97,17.7,109)\\ \hline
(300,600,1500)& (4.4,17.7,109)\\ \hline
(350,600,1500)& (6.05,17.7,109)\\ \hline
(400,600,1500)& (8,17.7,109)\\ \hline
(500,600,1500)& (12.3,17.7,109)\\ \hline
(600,600,1500)& (17.7,17.7,109)\\ \hline
\end{tabular}
\end{center}
\caption{ Eigen values of singlet fermion  masses for
  different allowed $N-S$ mixing matrix elements and for NH, IH, and QD
  patterns of light neutrino masses}   
\label{tableMms}
\end{table*}
\par
\begin{figure}[htbp]
\includegraphics[width=4cm,height=6cm, angle=-90]{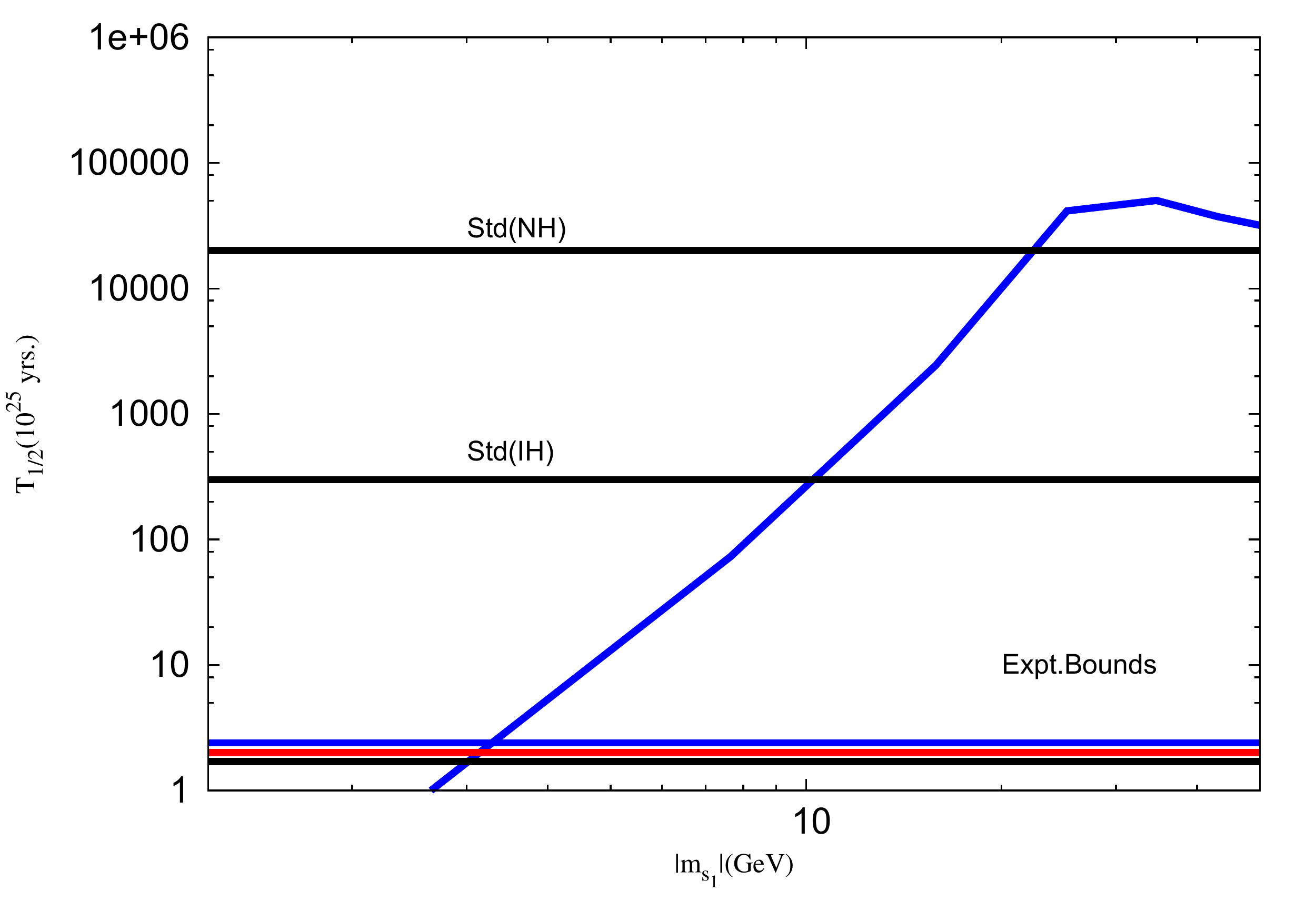}
\includegraphics[width=4cm,height=6cm, angle=-90]{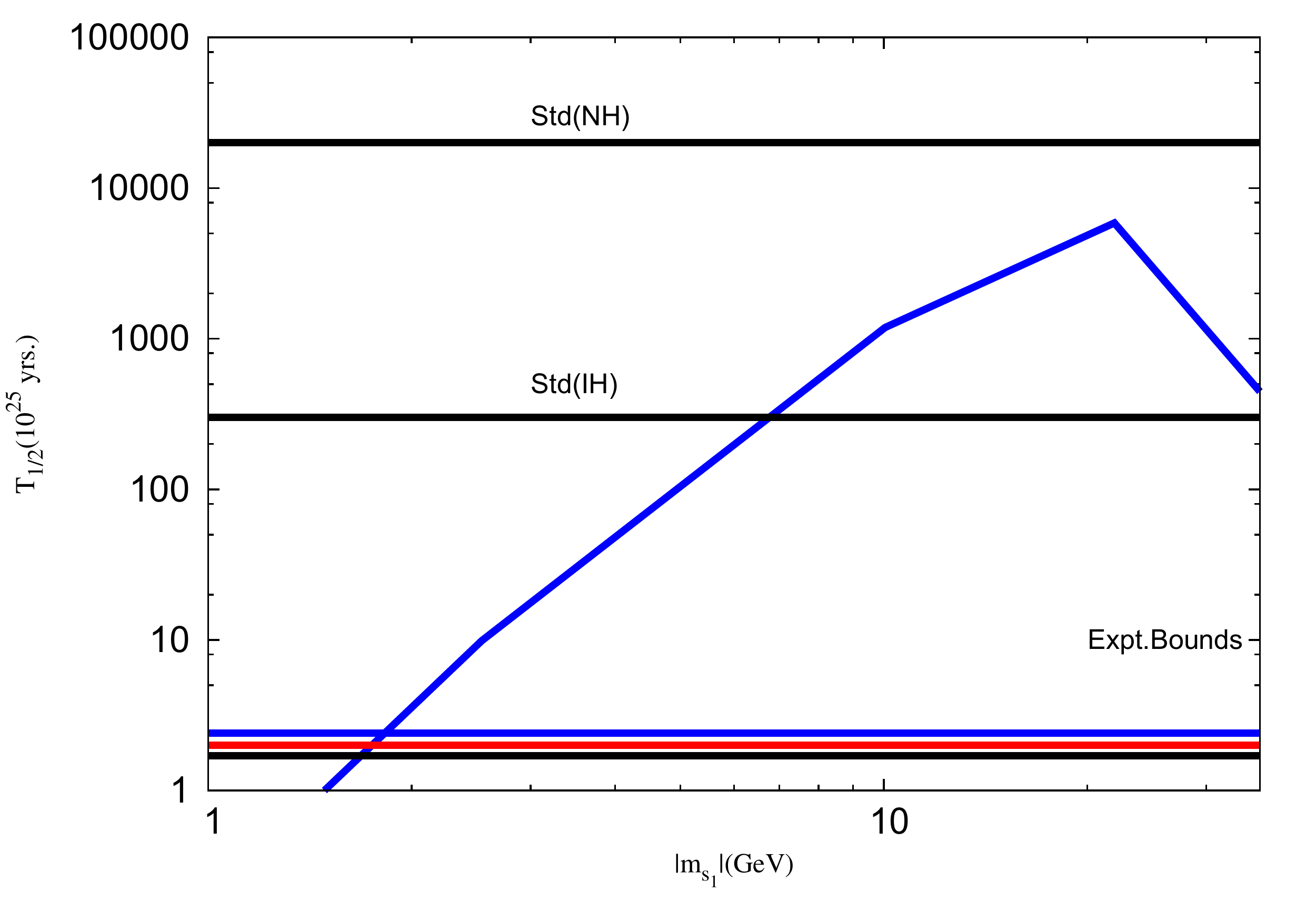}
\caption{Variation of half-life of $0\nu\beta\beta$ decay with the
sterile neutrino mass for NH(left) and IH(right) 
patterns of light active neutrino masses for $|p|=190$ {MeV} .}
\label{nrhs3}
\end{figure}

\noindent{\bf A. Half-life in the NH and IH cases}\\
We have computed the half-life for  NH and IH patterns of active neutrino masses,
taking the contributions of singlet fermion as well as  light
active neutrino exchanges. This is shown in the upper panel for NH case
and in the lower panel for IH case in Fig.\ref{nrhs3}.
 Taking both $X$ term and $Y$ term in eq.(\ref{XYhalfbb}),
  we find that for smaller value of  $m_{S_{1}}$, the contribution due
  to sterile neutrino is dominated for both NH and IH cases.
 But with the increase in the value of $m_{S_{1}}$, the half-lfe
 increases showing its decreasing strength. The predicted half-life
 curve saturates the experimental data at  $m_{S_{1}}\simeq 3$ GeV and
 $m_{S_{1}} \simeq 2$ GeV for the NH and the IH cases, respectively. The
 interesting predictions are that if the lightest sterile neutrino
 mass satisfies the bound  $m_{S_{1}}\le 3 GeV$, then the
 $0\nu\beta\beta$ decay should be detected with half-life close to the
 current experimental bound even if the light neutrino masses have NH
 pattern of masses. Similarly the corresponding bound for the IH case
 is  $m_{S_{1}}\le 2$ GeV. But in a recent paper\cite{pas} which has
 inverse seesaw dominant neutrino mass, the corresponding bound for the NH and IH case is
 $m_{S_{1}}\le 14$ GeV.
\vspace{0.2 cm} 
\par\noindent{\bf B. Lifetime prediction with QD neutrino masses.}\\
 
\begin{figure}[htbp]
\includegraphics[width=4cm, height=6cm,angle=-90]{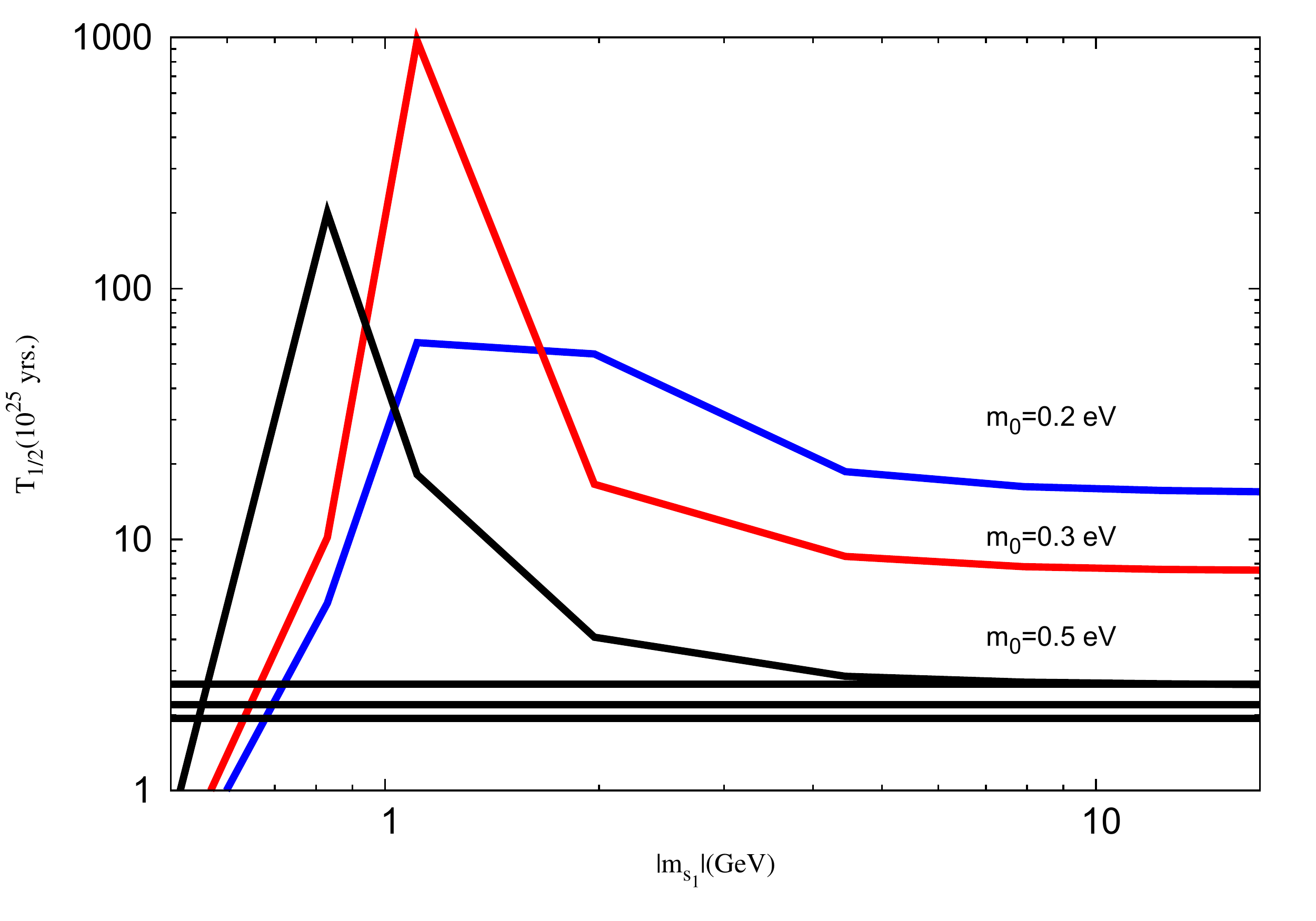}
\includegraphics[width=4cm, height=6cm,angle=-90]{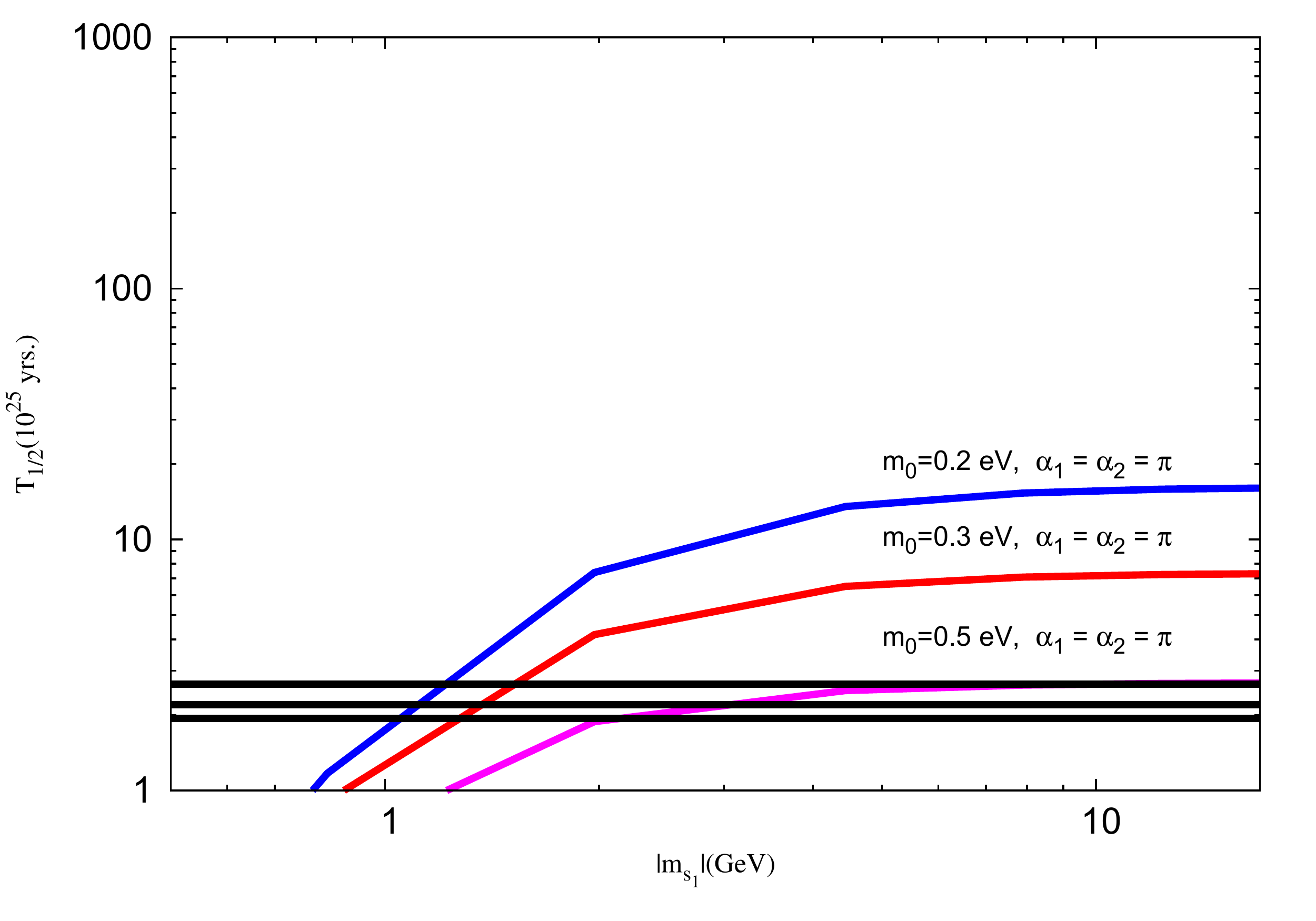}
\caption{Variation of half-life of $0\nu\beta\beta$ decay with the
  mass of the lightest singlet fermion for
 QD light neutrinos including one Dirac phase (left curve) and  one
 Dirac phase and two Majorana phases (right curve)
 .}
\label{chlsqd}
\end{figure}

  For QD masses of light active neutrinos, we considered the  $X$ term
 and $Y$ term of eq.(\ref{XYhalfbb}) i.e including both the sterile
 neutrino exchange  and light neutrino exchange contributions.
 For the light-neutrino effective mass parameter occuring in $Y$, we
 have considered three different cases with common light-neutrino 
 mass values $m_{0}=0.2
 {\rm eV}, 0.3 {\rm eV}$, and $0.5 {\rm eV}$  resulting in three
 different curves shown in the upper- and the lower- panels of
 Fig. \ref{chlsqd}. In the upper-panel, only the experimentally
 determined Dirac phase $\delta=0.8\pi$ has been included in the PMNS
 mixing matrix for light QD neutrinos while ignoring the two Majorana
 phases($\alpha_1=\alpha_2=0$). In the lower-panel while keeping
 $\delta=0.8\pi$ for all the three curves, the Majorana phases have
 been chosen as indicated against each of them.
As the sterile neutrino exchange amplitude given in eq.(\ref{effmassparanus2}) is
 inversely proportional to the eigen value of the corresponding
 sterile neutrino mass $m_{S_i}$, even in the quasi-degenerate case
 this contribution is expected to dominate for allowed small values of
 $m_{S_i}$. This fact is reflected in
 both the figures given in Fig.\ref{chlsqd}. When Majorana phases are
 ignored, this dominance gives half-life less than the current bounds
 for $m_{S_1} < 0.5$ GeV when $m_0=0.5$ eV, but for $m_{S_1} < 0.7$ GeV
 when $m_0=0.2-0.3$ eV. When Majorana phases are included preventing
 cancellation between the two contributions, these crossing points are
 changed  to $m_{S_1} < 0.7$ GeV when $m_0=0.3$ eV, but  $m_{S_1} < 1.0$ GeV
 when $m_0=0.2-0.5$ eV. Repeating the same procedure for
 ref. \cite{pas} which 
is based upon inverse seesaw dominance, the corresponding bound for the QD case is
 $m_{S_{1}}\le 12.5$ GeV.

In the present case, the peaks in the half-life prediction appear because of cancellation between the two effective
 parameters. Inclusion of Majorana phases annuls cancellation
 resulting in constructive addition of the two effective mass
 parameters and reduced values of half-life accessible to ongoing searches. 
For larger values of $m_{S_1}>> 20$ GeV, the sterile neutrino
contribution to $0\nu\beta\beta$ amplitude becomes negligible and the
usual contributions due to light quasi-degenerate neutrinos are recovered. \\
\section{Brief discussion on other aspects and
  leptogenesis}\label{secrv}
Here we discuss briefly constraints imposed on the model by
electroweak precision observables and predictions on the order of
magnitude of baryon asymmetry of the universe through resonant leptogenesis\cite{pilafts:2003}. We
also point out occurence of small $Z-Z'$ mixings while indicating
briefly a possible application for dilepton production. Since
details of analyses and predictions on these aspects are beyond the
scope of this paper, they will be presented elsewhere \cite{np}  
\par\noindent{(a)\bf Electroweak precision observables and other constraints}\\
We have shown that dominant contributions to $0\nu\beta\beta$ decay
are possible for the first generation sterile
neutrino masses ${\hat m}_{S_1} \sim {\mathcal O} (1)$ GeV. For larger values of this
mass  ${\hat m}_{S_1} \sim 5-10$ GeV partial cancellation between effective
mass parameters due to light neutrino and sterile neutrino exchanges
occurs depending upon choices of different Majorana phases. Different
lighter sterile mass eigen values relevant for $0\nu\beta\beta$ decay
are shown in Table \ref{tableMms} in the NH, IH, and QD cases. It is
pertinent to discuss influence of these lighter masses on the precision
electroweak observables.\\
For choices of parameters permitted by observable LFV and/or dominant LNV, the
sterile fermion masses of the first two generations could be ${\hat
  m}_{s_i} < 45$ GeV ,$i=1,2$ whereas in the absence of dominant LNV decay,
the mass eigen values could be even larger ${\hat
  m}_{s_i} \simeq 500$ GeV. When they are in the range of $1-45$ GeV,
we have estimated the corresponding corrections on electroweak observables. The $\nu-S$
mixing ${\mathcal V}^{\nu S}=({M_D\over M})_{\nu S}$ is well determined in
our model and all the relevant $\nu-S$ mixings are easily deduced
using eq.(\ref{MDatMR0}) and eq.(\ref{costr}).
In the allowed kinematical region, we have estimated the partial decay
widths,
\begin{equation}
 \Gamma (Z \to S_iS_i)=\Gamma^{\nu{\bar {\nu}}}_{Z} [\sum^{}_{\alpha}
  |\left(\mathcal{V}^{\nu S}_{\alpha,i}\right)|^4]\,(i=1,2), \label{gammazii}    
\end{equation}
where the standard value $\Gamma^{\nu{\bar {\nu}}}_{Z}=0.17$ GeV and
$\mathcal{V}^{\nu S}_{\alpha ,i}=(M_D/M)_{\alpha,i}$ with
 $\alpha=\nu_e,\nu_{\mu},\nu_{\tau}$ and $i=1,2,3$. We then obtain $\Gamma (Z
\to S_1S_1)= 1.2\times 10^{-14}$ GeV for NH, IH, and QD cases, and $\Gamma (Z
\to S_2S_2)=6.6\times 10^{-11}$ GeV for QD case only.  Similarly we have estimated the partial decay width
\begin{equation}
\Gamma(W \to l S_i)=\Gamma^{l {\nu}}_{W} [\sum^{}_{\alpha} |\left(\mathcal{V}^{\nu
  S}_{\alpha ,i}\right)|^2]\,(i=1,2),\label{gammawii}           
\end{equation}
and obtained $\Gamma(W \to e S_1) \simeq \Gamma(W \to e S_2) =3.5\times
 10^{-9}$ GeV,  $\Gamma(W \to \mu S_1) \simeq \Gamma(W \to \mu
S_2)= 1.8\times 10^{-7}$ GeV, and $\Gamma(W \to \tau S_1) \simeq \Gamma(W
\to \tau S_2) =  1.0\times 10^{-5}$ GeV. These and other related
estimations cause negligible effects on electroweak precision
observables \cite{ewo} primarily because of small $\nu-S$ mixings
determined by the model analyses. 
In addition to these insignificant tree level corrections, new
physics effects may affect the electroweak observables indirectly via
oblique corrections through loops leading corrections to the
Peskin-Takeuchi $S, T, U$ parameters
\cite{peskin:1990,appelquist:2003}. Although the computation of these loop effects are beyond the
scope of the present paper, it may be interesting to estimate how the new
fermions through their small mixings with active neutrinos may affect
the leptonic and the invisible decay widths of the
Z-boson, the W-mass, and other observables \cite{np}.      
 
In this model the neutral generator corresponding to heavy
$Z^{\prime}$ 
 is a linear combinations of $U(1)_R$
and $U(1)_{B-L}$ generators while the other orthogonal combination is
the $U(1)_Y$ generator of the SM \cite{mkpzp,langackerzp}.  The $Z-Z^{\prime}$ mixing in such theories
is computed through the generalised formula
$\tan^2{\theta_{zz^{\prime}}}=\frac{M_0^2-M_Z^2}
{M_{Z^{\prime}}^2-M_0^2}$ where $M_0= \frac{M_W}{{\sqrt \rho_0}\cos
  \theta_W}$.
In our model since the LH
triplet $\Delta_L(3,-1,1)$ has a very small VEV $v_L=0.1-0.5$ eV $<<
V_{\rm ew}$, the
model is consistent with the tree level value  
$\rho_0=1$. The radiative corrections due to the $125$ GeV Higgs of the SM and the
top quark yield $\rho\simeq 1.009$ \cite{jegerlehner:2012}.
The new neutral gauge boson $Z^{\prime}$ in principle may have
additional influence on the electroweak precision parameters as well as the
$Z-$pole parameters if $M_{Z^{\prime}} << {\cal O}(1)$ TeV \cite{langackerzp,erlerlanga:2009}.
The most recent LHC data has given the
lower bound   $M_{Z^{\prime}} \ge 1.6$ TeV \cite{LHC:Zp}.
 Since our model is
based on extended seesaw mechanism, we require $V_R >> V_{ew}=246$ GeV
and this implies $M_{Z^{\prime}}>> M_Z$ but accessible to LHC. Under
this constraint $M_{Z^{\prime}} \sim {\cal O} (5-10)$ TeV are the most
suitable predictions of both the models discussed in this work.
 As some examples, using such values of  $M_{Z^{\prime}}$ 
and the most recently reported values from Particle Data Group \cite{PDG:2015} of $\sin^2\theta_W=0.23126\pm 0.00005$,
 $M_W=80.385\pm0.015$GeV, $M_Z=91.1876\pm 0.0021$ GeV, $\rho_0=1.01$, we obtain 
$\theta_{zz^{\prime}}= 0.00131\pm 0.0003,\,\, 0.0005\pm 0.00012,\,\, 0.0003\pm
0.00008$,\,\, and $0.0002\pm 0.00006$ for
$M_{Z^{\prime}}= 2.0 {\rm TeV},\,\,5.0 {\rm TeV},\,\, 7.5 {\rm TeV}$,\,\, and $10 {\rm TeV}$
,respectively. Because of the smallness of the values,
 these mixings are consistent with the electroweak
precision observables including the $Z-$ pole data
\cite{langackerzp,erlerlanga:2009,montero:2010}. Some of these  masses may be also in the
accessible range of the ILC \cite{osland:2012}. Details of experimental
constraints on $Z-Z^{\prime}$ mixings as a function of $Z^{\prime}$ masses would
be investigated elsewhere \cite{np}. 
\par\noindent{\bf (b)Possibility of dilepton signals at LHC}\\
In both the models considered in this work, there are two types of heavy Majorana neutrinos: (i) the
RH neutrinos with masses $M_{N_i}\ge {\cal O} (1-10)$ TeV, (ii) some of the
three sterile neutrinos with masses ${\hat m}_{S-i}<<
M_{N_i}$. In principle both of these classes of fermions are capable
of contributing to dilepton production at LHC through the sub-processes
$pp\to W^{\pm}_L \to l^{\pm}l^{\pm}jjX$ where, for example, the $W_L^+$ produced from
$pp$ collision gives rise to a charged lepton $l^+$ and a $N_i$ or $S_i$ in the
first step by virtue of the latters' mixing with the charged leptons
given in eq.(\ref{ModCC}).
 The particle $N_i$ or $S_i$ can then produce a second charged
lepton of the same sign and a $W^-_L$ boson that is capable of giving rise
to two jets. It is interesting to note that our model predicts a rich
structure of
like sign dilepton production through the mediation of $N_i$ or $S_i$,
or both. 
From details of model parametrisations discussed in Sec.3-Sec.5, we
have found the corresponding mixing matrices with charged leptons
defined through eq.(\ref{ModCC}) discussed in Sec.5.
We have estimated the elements ${\cal V}^{\nu N}_{e 1}\simeq -0.0000727+i0.000203$ and ${\cal
    V}^{\nu N}_{\mu 2}\simeq 0.000813-i0.001148$
which would
  contribute to the production cross sections of $pp \to e^{\pm}e^{\pm}jjX$ and $pp \to \mu^{\pm}\mu^{\pm}
  jjX$ by the exchange of RH neutrinos, the cross sections being
  proportional to the modulus squares of these mixings. Similarly we have 
 ${\cal V}^{\nu S}_{\mu 2}\simeq 0.0003191$ which can also contribute
    to production process $pp\to \mu^{\pm}\mu^{\pm}jjX$ by the exchange
    of the second sterile
      neutrino mass eigen state. The first sterile neutrino
      is too light
      to mediate the dilepton production process. Thus the LHC
      evidence of dilepton production signals, may indicate the
      presence of heavy Majorana neutrinos \cite{almeida:2000}.  Details of predictions will
      be reported elsewhere \cite{np}.
\par
\begin{figure}[htbp]
\begin{center}
\includegraphics[width=4 cm,height=3cm]{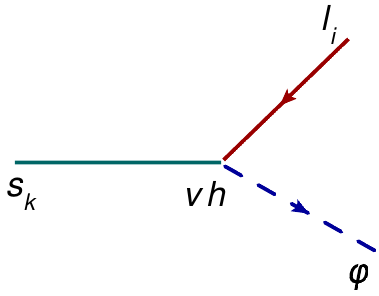}
\includegraphics[width=4 cm,height=3 cm]{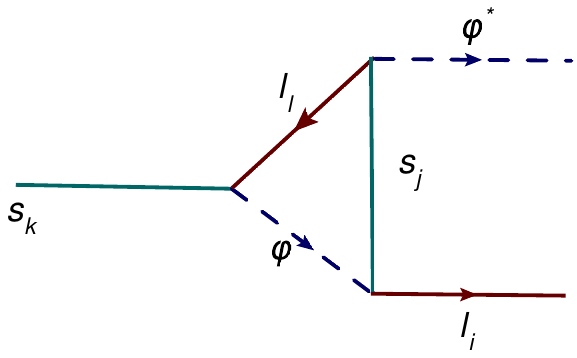}
\includegraphics[width=4 cm,height=3 cm]{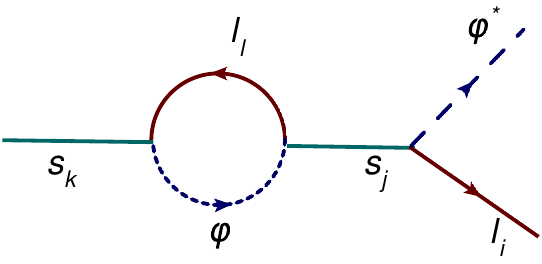}
\caption{Tree and one-loop diagrams for the $S_{k}$ decay contributing to the CP-asymmetry. All fermion-Higgs couplings in the diagrams are of the
form $Vh$ where $h= N-l-\Phi$ Yukawa coupling and $V\simeq M/M_N$.}
\label{fig:Fvertex}
\end{center}
 \end{figure}

\noindent{\bf (c)Leptogenesis}\\
This model may have a wider range of possibilities for leptogenesis via
decays of Higgs triplets \cite{HamGoran}, or through the decays of LHC
scale Majorana fermions $N$ or $S$. Although rigorous estimation
including solutions of Boltzmann equations is beyond the scope of this
work which will be addressed elsewhere\cite{np}, we discuss here
briefly only a plausible case with a very
 approximate estimation of the CP
asymmetry parameter and the order of magnitude of the baryon asymmetry 
through the decays of two nearly degenerate Majorana masses of
sterile neutrinos. For resonant leptogenesis through the decays of a pair of
quasi-degenerate RH neutrinos, relevant formulas for
CP-asymmetry and baryon asymmetry have been suggested in \cite{pilafts:2003}.
 Noting that ${\hat m}_{S_1}\sim {\mathcal O}(1)$ GeV is
important for dominant contribution to $0\nu\beta\beta$ decay and the
$N-S$ mixing matrix elements $M_2
\sim M_3 \simeq {\mathcal O} (1) $ TeV are capable of predicting
experimentally accessible LFV decays in our model, we choose an interesting region of the parameter
space $M \simeq {\rm diag.} (146,3500,3500)$ GeV in the
quasi-degenerate  case of 
$S_2$ and $S_3$. Then 
using the $G_{2113}$ breaking VEV $V_R \simeq {\mathcal O }(10)$ TeV,
the results of Sec. 3.3 in the QD case of active neutrinos, and eq.(\ref{massmatrices})
 we obtain
\bea 
{\hat m}_{S_i} &=& {\rm diag.} (1.0, 595.864.., 595.864..){\rm
  GeV}.\label{msi}
\eea 
where ellipses on the RHS indicate higher degree of quasi-degeneracy
of the two masses the model tolerates.
In order to estimate lepton asymmetry caused by the decay of heavy
sterile fermions ${\hat S}_k (k=2,3)$ via their mixing with the heavier RH
neutrinos, the corresponding 
 Feynmann diagrams at the tree and one-loop levels, including the
 vertex and self energy diagrams, are
 shown in Fig. \ref{fig:Fvertex}.
The fermion Higgs coupling in all the diagrams is
 $Vh$ instead of the standard Higgs-Yukawa coupling 
$h=M_D/ V_{\rm  wk}$ where  ${\mathcal V}\simeq{M/M_N}$,
$M_D$ is given in eq.(\ref{MDatMR0}), and $V_{\rm  wk}\simeq 174$ GeV.    
The widths of these sterile fermion are
 $\Gamma_{S_2}\simeq 16.3\,\, {\rm keV}$ and $\Gamma_{S_3}\simeq 14.0\,\, {\rm MeV}$.
In order to exploit quasidegeneracy of the second and the third
generation fermions in resonant leptogenesis, we use the formula for CP asymmetry generated due
to interference between the tree and the self energy graphs \cite{pilafts:2003}, 
\bea
\varepsilon_{S_k}&=& \sum_j\frac{{\cal I}m[(y^{\dagger}y)_{kj}^2]}{
  |y^{\dagger}y|_{jj}|y^{\dagger}y|_{kk}}R  \nonumber\\
R&=&\frac{({\hat m}_{S_i}^2-{\hat m}_{S_j}^2){\hat
    m}_{S_i}\Gamma_{S_j}}{({\hat m}_{S_i}^2-{\hat m}_{S_j}^2)^2
+{\hat m}_{S_i}^2\Gamma_{S_j}^2} \,,
\label{epsN}
\eea
where $y= M/M_{N}h$, $h=M_D/V_{\rm wk}$ ,and $V_{\rm wk}\simeq
174$ GeV.
For computation of the baryon asymmetry $Y_B$ with a given washout
factor $K$, we have also utilised the
suggested formula\cite{pilafts:2003} 
\bea
Y_B &\simeq& \frac{\varepsilon_{S_k}}{200 K_k},\nonumber\\ 
K_k &=&\frac{\Gamma_{S_k}}{H({\hat m}_{S_k})}, \label{bau}  
\eea
$H({\hat m}_{S_k})$ being the Hubble parameter at temperature ${\hat m}_{S_k}$.
As in TeV scale leptogenesis models, here also we encounter large
wash-out factors which, in some cases, tend to damp out the baryon
asymmetry generation. However it has been shown 
\cite{blhammi} that all the $\Delta L =2$ processes, $l\Phi \to {\bar
  l}\Phi^{\dagger}$ expected to cause the most dominant washouts are 
substantially depleted for the heavier quasidegenerate Majorana masses of the decaying
fermions. The depletion factor is
proportional to $\delta_i^2$ leading to an effective washout factor
$K_i^{\rm eff}$ that replaces $K_i$ for the $i-$th decaying Majorana fermion
\bea
\delta_i&=&\frac{|{\hat m}_{S_i}-{\hat m}_{S_j}|}{\Gamma_{S_i}}(i\neq j), \nonumber\\
K_i^{\rm eff}&\simeq &\delta_i^2 K_i.\label{keff}
\eea 
We find sizeable baryon asymmetry in the following two cases:
(i)In the case of finite perturbation theory, the ${\hat m}_{S_j}^2\Gamma_{S_j}^2$ term
in the denominator of $R$ has been noted to be absent\cite{pilafts:2003} leading to a singular term in the CP-asymmetry. 
(ii)In the limit when $|{\hat m}_{S_i}-{\hat m}_{S_j}|<< \Gamma_{S_j}/2$, 
$R=2{({\hat m}_{S_i}-{\hat m}_{S_j})\over\Gamma_{S_j}}$.  
\par\noindent{(i)Finite perturbation theory}\\
\bea
R&=& {1\over 2}\frac{\Gamma_j}{({\hat m}_{S_j}-{\hat
    m}_{S_k})},\nonumber\\
\delta_{jk}&=& (1-{{\hat m}_{S_j}\over {\hat m}_{S_k}}),\nonumber\\
\varepsilon_{S_k}&=& \sum_j\frac{{\cal I}m[(y^{\dagger}y)_{kj}^2]}{16\pi
  |y^{\dagger}y|_{kk}\delta_{jk}}.\label{FPT}
\eea
Similar formulas have been used by a number of authors in the case of
decays of quasi-degenerate RH neutrinos \cite{ahn} and, specifically, in
the context of $SO(10)$\cite{pilafts:1997}.  
For the decay of $S_2$ for which $K_2= 2.7\times 10^7$,  using $({\hat m}_{S_2}-{\hat m}_{S_3})\simeq
2\times 10^{-7}$ GeV, we obtain
\bea
\varepsilon_{S_2}&=& 0.824,\nonumber\\
Y_B&=& 1.5\times 10^{-10}.\label{bau2f}
\eea
The fine tuning in the quasidegenerate masses can be reduced by one order if
we use the effective wash out factor. For example using $({\hat
  m}_{S_2}-{\hat m}_{S_3})\simeq 1.35\times 10^{-6}$ GeV, we get
$\delta_2=0.008$ leading to 
\bea
\varepsilon_{S_2}&=& 0.0357,\nonumber\\
K_2^{\rm eff}&=&1.92\times 10^5,\nonumber\\
Y_B&=& 9.3\times 10^{-10}.\label{bau2f}
\eea

For the decay of $S_3$ for which $K_3= 2.4\times 10^{10}$,  using $({\hat m}_{S_2}-{\hat m}_{S_3})\simeq
 10^{-6}$ GeV, we obtain $\delta_3^2 \simeq 5\times 10^{-8}$,
  leading to
 \bea
\varepsilon_{S_3}&=& 8\times 10^{-5},\nonumber\\
 K_3^{\rm eff}&=&575.4,\nonumber\\
Y_B&=& 7.3\times 10^{-10}.\label{bau3f}
\eea  
\par\noindent{(ii)Larger width limit :$\Gamma_k >2|({\hat m}_{S_k}-{\hat m}_{S_j})|$}\\
\bea
R &\simeq& \frac{2({\hat m}_{S_k}- {\hat
    m}_{S_j})}{\Gamma_k},\nonumber\\
&&\times \left [1+\frac{4({\hat m}_{S_k} - {\hat
    m}_{S_j})^2}{\Gamma_k^2}\right]^{-1},\nonumber\\
\varepsilon_{S_k}&=& \sum_j\frac{{\cal I}m[(y^{\dagger}y)_{kj}^2]}{8\pi
  |y^{\dagger}y|_{kk}\Gamma_k}{\hat m}_{S_k}R. \label{LW}
\eea
This case can be more efficiently implemented for $S_3$ decay which has
$\Gamma_{S_3}\simeq 14$ MeV, and
 $K_3= 2.4\times 10^{11}$. In this case the depletion in $K_3$ is quite effective. Using $({\hat m}_{S_2}-{\hat m}_{S_3})\simeq
 10^{-6}$ GeV, we obtain $\delta_3^2 \simeq 5\times 10^{-11}$,
 leading to
 \bea
\varepsilon_{S_3}&=& 3\times 10^{-7},\nonumber\\
K_3^{\rm eff}&=&10.3,\nonumber\\
Y_B&=& 1.1\times 10^{-10}.\label{bau3j}
\eea  
Thus we have shown very approximately that the model may be capable of accommodating the order of magnitude of baryon asymmetry of the
universe that requires fine tuning of the mass difference of the
two sterile neutrino in the range $10^{-6}-10^{-7}$ GeV. 
In a separate paper we plan to look into improvement in these approximate solutions and other
possible channels of leptogenesis including the impact of the present
model on electroweak precision observables and detection possibilities of
RH neutrinos, $S-$ fermions, and the $Z^{\prime}$ at collider energies
such as LHC and ILC\cite{np}.

\section{Summary and conclusion}
In this work we have investigated the prospect of having a new type-II seesaw
dominated neutrino mass generation mechanism  in non-SUSY $SO(10)$ GUT 
by a novel procedure by introducing one additional singlet fermion per
generation. Following the popular view that the only meaningful fermion masses in the
Lagrangian must have dynamical origins, and taking the non-dynamical singlet fermion
mass $\mu_S$ to be negligible, one of the models (Model-I)
discussed
is found to exhibit type-II seesaw dominance and it predicts TeV scale
$Z^{\prime}$ boson accessible to LHC without any drastic fine tuning
in the corresponding Yukawa sector. For Model-II the desired type-II
seesaw dominance requires an additional fine tuning upto one part in a
million. The would be dominant type-I
seesaw contribution to neutrino masses in both models cancels out. The induced contribution to the $\nu-S$ mixing
mass term $M_L$  is shown to be damped out because of the GUT-scale mass
of the LH doublet in ${16}_H$ that renders the linear seesaw
contribution to light neutrino masses naturally
negligible in the Model-I, although in Model-II it needs additional
fine tuning.
In spite of the high
values of the type-II seesaw scale $M_{\Delta_L}\simeq 10^8- 10^9$ GeV $>> M_Z$, the
models predict new dominant contributions to $0\nu\beta\beta$ decay in
the $W_L-W_L$ channel mediated by sterile neutrinos which acquire
Majorana masses. The predicted LFV decay branching ratios for $\mu \to
e \gamma$,  $\tau \to \mu \gamma$, and $\tau \to
e \gamma$,  are found to be accessible to ongoing and planned
experiments.  We discuss the impact on the resultant
effective mass parameter and   $0\nu\beta\beta$ half-life showing
cancellation between light-neutrino exchange and sterile neutrino
exchange contributions. The cancellation occurs because of the
opposite signatures of the two effective mass parameters due to light
neutrino exchange and the sterile neutrino exchange when effects of
Majorana phases are ignored.     
We derive an analytic formula for the half-life
of $0\nu\beta\beta$ decay as a function of singlet fermion masses
which predicts a lower bound on the lightest sterile neutrino mass
eigen value from the current experimental data on lower bounds.
 We find that the half-life close to the current
lower bound or even lower can be easily accommodated even with NH or IH
patterns of light neutrino masses . We find that the QD nature of light neutrino
masses is not a necessary criteria to satisfy existing lower bounds on
the half life estimated by different experimental groups. Even if the light active neutrino
masses are NH or IH, a half-life prediction $T_{1/2}\simeq (2-5)\times 10^{25}$ yrs
is realizable if the lightest sterile neutrino mass $m_{S_1}\simeq
2-3$ GeV. Depending upon the common mass of the light QD neutrinos,
the model also predicts lifetime  $T_{1/2} \leq 2 \times 10^{25}$ yrs
for $m_{S_1}\leq (0.5-1.0)$ GeV. Large cancellation between the two
contributions is found to occur in the quasidegenerate case of light active
neutrinos in the regions of sterile neutrino mass $m_{S_1}\simeq 2-8$
GeV. The bounds obtained in the sterile neutrino mass  in these
type-II seesaw dominant models are significantly smaller than that of the 
bounds obtained  in the inverse seeseaw model \cite{pas}.
As the sterile neutrino contribution to the $0\nu 2\beta$ decay is
inversely proportional to the corresponding mass eigen values, the
smallness of the lightest mass eigen values causes dominant
contributions compared to those by light neutrinos in NH, IH, and QD
cases. For the same reason the new contributions are damped out for
large sterile neutrino mass eigen values.  
Because of the underlying type-II seesaw formula for neutrino masses,
heavy RH neutrino masses in the
range ${\mathcal O}(100)$ GeV-${\mathcal O}(10000)$ GeV and with
specified 
heavy-light neutrino mixings are also predicted 
which can be  testified at the LHC and
future high energy accelerators.  
  The proton lifetime predictions for $p\to e^+\pi^0$ for some regions
  of the parameter space
 are also accessible to ongoing  experimental searches especially for
intermediate mass values of the color octet scalar 
which has been found to be necessary for gauge coupling unification.
 Further we have verified that the lighter $S_1$ or $S_2$
states in the models have negligible effects on values of electroweak
precision observables at the tree level although loop effects through
Peskin-Takeuchi parameters $S, T, U$ will be
investigated elsewhere. Approximate estimations show occurence of
small $Z-Z'$ mixings apparently consistent with $Z-pole$ and non-$Z-pole$ data.
 The possibility
of dilepton signals at LHC in the $W_L-W_L$ channel is briefly noted
in both the models while an approximate estimation indicates possibility of 
baryon asymmetry generation through leptogenesis due to decay of
quasidegenerate  sterile Majorana fermions at the TeV scale. The details and rigorous
estimations on dilepton signals, leptogenesis, estimation of $S,T, U$
parameters, and the impact of $Z-Z'$ mixings on $Z-pole$ and non-$Z$
pole data including electroweak precision observables
are currently under investigation and would be reported separately\cite{np}. 

\noindent{\large \bf Acknowledgment}\\
M. K. P. thanks Thomas Hambye for discussion and the Department of Science and Technology, Govt. of
India for the research project, SB/S2/HEP-011/2013. B. N. thanks Siksha 'O' Anusandhan
University for a research fellowship.
\subsection{Appendix A}
\noindent{\large \bf { Beta function coefficients for RG evolution of
    gauge couplings}}\\
The renormalisation group equations for gauge couplings are 
\begin{equation}
\mu{\partial g_i\over {\partial
      \mu}}=\frac{a_i}{16\pi^2}g_i^3+\frac{1}{(16\pi^2)^2}\sum_jb_{ij}g_i^3g_j^2,\label{RGgi}
\end{equation}
where $a_i$($b_{ij}$) are one-(two-)loop beta function
coefficients. Their values for the Model-I and Model-II are given in
Table \ref{tabgi}.
\vspace{0.1cm}
\begin{table}
\begin{tabular}{|c|c|c|}
\hline
 $Symmetry$&${a_i}$&${b_{ij}}$\\
         &&(GeV)\\ \hline
$G_{213}$&$\begin{pmatrix}-19/6,41/10,-7\end{pmatrix}$&$\begin{pmatrix}
199/50,27/10,44/5\\9/10,35/6,12\\11/10,9/2,-26\end{pmatrix}$ \\ \hline
$G_{2113}$&$\begin{pmatrix}-3,57/12,37/8,-7\end{pmatrix}$&$\begin{pmatrix}8,1,3/2,12\\3/2,33/57,63/8,12\\9/2,63/8,209/16,4\\9/2,3/2,1/2,26\end{pmatrix}$ \\ \hline
$G_{2213}$&$\begin{pmatrix}-2,-3/2,29/4,-7\end{pmatrix}$&$\begin{pmatrix}31,6,39/2,12\\6,115/6,3/2,12\\81/2,6,181/8,4\\9/2,9/2,1/2,-26\end{pmatrix}$ \\ \hline
$G_{2213D}$&$\begin{pmatrix}-3/2,-3/2,15/2,-7\end{pmatrix}$&$\begin{pmatrix}319/6,6,57/4,12\\6,319/6,57/4,12\\171/4,171/4,239/4,4\\9/2,9/2,1/2,-26\end{pmatrix}$\\ \hline
\end{tabular}
\caption{One-loop and two-loop beta function coefficients for gauge coupling
  evolutions described in the text taking the second
  Higgs doublet mass at $1$ TeV} 
\label{tabgi}
\end{table}    
\vspace {3cm}
\subsection{Appendix B}
\noindent{\large \bf { Block diagonalisation and determination of $\cal{M}_{\nu}$}}\\
In this section we discuss the various steps of block diagonalisation in order to calculate the light neutrino mass , sterile neutrino mass and right-handed neutrino mass and their mixings.
The complete $9\times9$ mass matrix in the flavor basis  $\{\nu_L, S_L, N^C_R\}$ is 
\bea
\mathcal{M}=
\bmt
m_{\nu}^{II}    &   M_L     &  M_D  \\
M_L^T &   0       &  M  \\
M_D^T &   M^T     &  M_N
\emt
\label{app:numass}
\eea ,
where $M_L=y_{\chi}v_{\chi_L}$, $M=y_{\chi}v_{\chi_R}$, $M_N=fv_R$\\
and $M_D$ is the Dirac neutrino mass matrix as discussed in Sec.4.\\
Assuming a generalized unitary transformation from mass basis to flavor basis, gives
\bea 
& &|\psi\rangle_{flavor}=\mathcal{V}\, |\psi\rangle_{mass} 
\eea
\label{aform1}
or\\
\bea
&\mbox{}&\,\bmt 
\nu_\alpha\\ S_\beta \\ N^C_\gamma
\emt
=
\bmt 
{\cal V}^{\nu\nu}_{\alpha i} & {\cal V}^{\nu{S}}_{\alpha j} & {\cal V}^{\nu {N}}_{\alpha k} \\
{\cal V}^{S\nu}_{\beta i} & {\cal V}^{SS}_{\beta j} & {\cal V}^{SN}_{\beta k} \\
{\cal V}^{N\nu}_{\gamma i} & {\cal V}^{NS}_{\gamma j} & {\cal V}^{NN}_{\gamma k} 
\emt
\bmt 
\hat{\nu}_i \\ \hat{S}_j \\ \hat{N}_k
\emt  
\eea
 \label{app:formmix}
\\
with\\
\bea
&\mbox{} &\,\mathcal{V}^\dagger \mathcal{M} \mathcal{V}^*
    =  \hat{\mathcal{M}}
	 = {\rm diag}\left({ \hat{\mathcal M}}_{\nu_i};{ \hat{\mathcal M}}_{{\cal S}_j};{ \hat{\mathcal M}}_{{\cal N}_k}\right)
	 \label{app:massdiag}
\eea
Here $\mathcal{M}_\nu$ 
is the $9\times 9$ neutral fermion mass matrix in flavor basis with $\alpha, \beta, \gamma$ running over three generations of light-neutrinos, 
sterile-neutrinos  and right handed heavy-neutrinos in  their respective flavor states and $\hat{\mathcal{M}}_\nu$ is 
the diagonal mass matrix with $(i,j,k=1,2,3)$ running over corresponding mass states .\\
In the first step of block diagonalisation, the full neutrino mass matrix is reduced to a block diagonal form $\hat{\mathcal{M}}_{\rm \tiny BD}$ and
in the second step we further block diagonalize to obtain the three matrices as three different block diagonal elements, $\mathcal{M}_{\rm \tiny BD}$= $diag({\mathcal M_{\nu}},{m_S},{m_N})$  
whose each diagonal element is a $3\times 3$ matrix.In our estimation, we have used the mass hierarchy $M_N > M \gg M_D, M_L, fv_L$. Finally in the third step we discuss complete diagonalization
to arrive at the physical masses and their mixings.
\subsubsection{Determination of $\mathcal{M}_{\rm \tiny BD}$}
With two unitary matrix transformations $\mathcal{Q}_1$ and $\mathcal{Q}_2$, \\
 \bea
\mathcal{Q}^\dagger \mathcal{M}_\nu  \mathcal{Q}^* = \hat{\mathcal{M}}_{\rm \tiny BD}, \mbox{}\quad 
\eea
\label{rr}
where \\
\bea
\mathcal{Q}=\mathcal{Q}_1\,\mathcal{Q}_2
\eea
i.e the product matrix $\mathcal{Q}=\mathcal{Q}_1\, 
\mathcal{Q}_2$ directly give  $\mathcal{M}_{\rm \tiny BD}$ from $\mathcal{M}_\nu$
Here $\hat{\mathcal{M}}_{\rm \tiny BD}$, and $\mathcal{M}_{\rm \tiny BD}$ are the intermediate block-diagonal, 
and full block-diagonal mass matrices, respectively,
\bea 
& &\hat{\mathcal{M}}_{\rm \tiny BD} =
\bmt
{\cal M}_{eff}&0\\
0& {m_N}
\emt
\eea
and
\bea
&\mbox{}&\, \mathcal{M}_{\rm \tiny BD}
= \bmt {\cal M_\nu}&0&0\\
0&{m_S}&0\\
0&0&{m_N}
\emt
\eea
\subsubsection{\bf Determination of $Q_1$}
In the leading order parametrization the standard form of 
$\mathcal{Q}_1$ is
\bea 
\mathcal{Q}_1=\bmt
1-\frac{1}{2}R^*R^T&R^*\\
-R^T&1-\frac{1}{2}R^TR^*
\emt\,, 
\eea 
 where $R$ is a $6\times 3$ dimensional matrix.
\be
R^\dagger =M_N^{-1}\left(M^T_D,M^T\right)=(K^T, J^T)
 \label{app:w1}
\ee
\bea
J=M{M_N}^{-1}  
K=M_DM_N^{-1} 
I=KJ^{-1}=M_DM^{-1}   \label{kk}    
\eea 
Therefore, the transformation matrix $\mathcal{Q}_1$ can be written purely in terms of dimensionless parameters $J$ and $K$
\bea 
\mathcal{Q}_1=\bmt
1-\frac{1}{2}KK^\dagger & -\frac{1}{2}KJ^\dagger & K \\
-\frac{1}{2}JK^\dagger & 1-\frac{1}{2}JJ^\dagger & J \\
-K^\dagger & -J^\dagger & 1-\frac{1}{2}(K^\dagger K+ J^\dagger J)
\emt
 \label{app:w1}
\eea
while the light and heavy mass matrices are
\bea
{\cal M}_{eff}&=&\bmt
fv_L&M_L\\
M_L^T&0
\emt -
\bmt 
M_DM_N^{-1}M_D^T&M_DM_N^{-1}M\\
M^TM_N^{-1}M^T_D& M^TM_N^{-1}M
\emt \\
{m_N}&=&M_N+..
\eea
Denoting 
\bea
{\cal M}_{eff}&=&\bmt
Z&B\\
C&D
\emt \label{lk}
\eea , 
\bea
Z=fv_L-M_DM_N^{-1}M_D^T ,
\eea
\bea
B=M_L-M_DM_N^{-1}M ,
\eea
\bea
C=M_L^T-M^TM_N^{-1}M^T_D ,
\eea
\bea
D=M^TM_N^{-1}M , \label{sk}
\eea       
\subsubsection{Determination of $\mathcal{Q}_2$}
The remaining mass matrix ${\cal M}_{eff}$ can be further
block diagonalized using another transformation matrix
\bea 
\mathcal{S}^\dagger \mathcal{M}_{\rm eff} \mathcal{S}^* 
      = \bmt 
        {\cal M_\nu} & 0 \\
         0     & {m_S}
        \emt \label{op}
\eea
such that in  eq.(\ref{rr})
\bea
\mathcal{Q}_2 =  \bmt
               \mathcal{S}&0\\
               0&1
                 \emt 
\eea
 \bea 
 S=\bmt 
1-\frac{1}{2} P^*P^T&P^*\\
-P^T&1-\frac{1}{2}P^TP^*
 \emt      \label{pp} 
\eea    
Using eq.(\ref{pp}) in eq.(\ref{op}) ,we get through eq.(\ref{lk})-eq.(\ref{sk}), 
\bea 
P^\dagger &=&(M^TM_N^{-1}M)^{-1}\left(M^TM_N^{-1}M^T_D-M^T_L\right)\nonumber \\
&=& M^{-1}M_D^T-M^{-1}M_NM^{-1}M_L
\eea 
where we have used $y_{\chi}$ to be symmetric.
leading to
  \begin{eqnarray}
 {\mathcal M_\nu} = m_{\nu}^{II} + \left(M_D M_N^{-1} M^T_D\right) \nonumber\\
-(M_D M_N^{-1} M^T_D )+ M_L(M^T M_N^{-1}M)^{-1} M_L^{T}\nonumber \\
 -M_L(M^T M_N^{-1}M)^{-1}(M^T M_N^{-1}M_D^T)\nonumber\\ -(M_D M_N^{-1} M) (M^T M_N^{-1}M)^{-1} M_L^{T},\nonumber\\
 {m_ {S}} =-MM_N^{-1}M^T+.... ,\nonumber\\ 
 \end{eqnarray}  
 The $3 \times 3$ block diagonal mixing matrix $\mathcal{Q}_2$ has the following form
\bea 
\mathcal{Q}_2 
=\bmt 
S & {\bf 0} \\
{\bf 0} & {\bf 1}
\emt = 
\bmt 
1-\frac{1}{2}II^\dagger &I & 0\\
-I^\dagger & 1-\frac{1}{2}I^\dagger I & 0 \\
0 & 0 & 1
\emt
 \label{app:w2}
\eea
 where we have used eq.(\ref{kk}) to define $I=KJ^{-1}=M_DM^{-1}$.\\
  \noindent{\bf Complete diagonalization and physical neutrino masses}\\
The $ 3\times 3$ block diagonal matrices $\cal{M}_\nu$, $m_{S}$ and $m_{N}$ can further be diagonalized 
to give physical masses for all neutral leptons by a $9\times 9$  unitary matrix $\mathcal{U}$ as
\bea
\mathcal{U}=\bmt U_\nu & 0 & 0 \cr 0 & U_S & 0 \cr 0 & 0 & U_N \emt.
\label{eq:mixb}
\eea
where the $ 3\times 3$  unitary matrices $U_\nu$, $U_{S}$ and $U_{N}$ satisfy
\begin{eqnarray}
U^\dagger_\nu\, {\cal M_\nu}\, U^*_{\nu}  &=& \hat{\cal M_\nu} = 
         \text{diag}\left({\cal M_\nu}_1,{\cal M_\nu}_2, {\cal M_\nu}_3\right)\, , \nonumber \\ 
U^\dagger_S\, {m_ S}\, U^*_{S}  &=& \hat{m_S}= 
         \text{diag}\left({m_S}_1, {m_S}_2, {m_S}_3\right)\, , \nonumber \\
U^\dagger_N\, {m_ N}\, U^*_{N}  &=& \hat{m_N} = 
         \text{diag}\left({m_N}_1, {m_N}_2, {m_N}_3\right)\,
 \label{app:unit}
\end{eqnarray}
\noindent
With this discussion, the complete mixing matrix is
\begin{eqnarray}
\mathcal{V}=\mathcal{Q} \cdot \mathcal{U} =
\left(\mathcal{Q}_{1}\cdot \mathcal{Q}_{2}\cdot \mathcal{U}\right) \nonumber \\
=
\bmt
1-\frac{1}{2}KK^\dagger & -\frac{1}{2}KJ^\dagger & K \\
-\frac{1}{2}JK^\dagger & 1-\frac{1}{2}JJ^\dagger & J \\
-K^\dagger & -J^\dagger & 1-\frac{1}{2}(K^\dagger K + J^\dagger J)
\emt\cdot\nonumber\\
\bmt
1-\frac{1}{2}II^\dagger & I & 0 \\
-I^\dagger  &   1-\frac{1}{2}I^\dagger I& 0 \\
0 & 0 & 1
\emt
\bmt 
U_\nu &0&0\\
0&U_{S}&0\\
0&0&U_{N}
\emt  \nonumber \\
=
\bmt
1-\frac{1}{2}II^\dagger & I-\frac{1}{2}KJ^\dagger &K\\
-I^\dagger & 1-\frac{1}{2}(I^\dagger I+JJ^\dagger) & J-\frac{1}{2}I^\dagger K\\
0&-J^\dagger&1-\frac{1}{2}J^\dagger J
\emt
\cdot
\bmt 
U_\nu &0&0\\
0&U_{S}&0\\
0&0&U_{N}
\emt 
 \label{app:mix-extended}
 \end{eqnarray}

\end{document}